\documentclass[%
preprint,
 amsmath,amssymb, aps,
]{revtex4-2}

\usepackage{graphicx}% Include figure files
\usepackage{dcolumn}% Align table columns on decimal point
\usepackage{bm}% bold math
\usepackage[dvipsnames]{xcolor}

\begin{document}

%\preprint{APS/123-QED}

\title{Cross-feeding yields high-dimensional chaos and coexistence of species beyond exclusion principle}% Force line breaks with \\
%\thanks{A footnote to the article title}%

\author{Takashi Shimada}
 \altaffiliation[Also at ]{Mathematics and Informatics Center, The University of Tokyo.}%Lines break automatically
 \email{takashi_shmd@gmail.com}
%\homepage{http://www.Second.institution.edu/~Charlie.Author}
\affiliation{%
 Department of Systems Innovation, Graduate School of Engineering, The University of Tokyo, 7-3-1 Hongo, Bunkyo-ku, Tokyo 113-8656 Japan
}%

%\collaboration{MUSO Collaboration}%\noaffiliation

\author{Kunihiko Kaneko}
\affiliation{Niels Bohr Institute, University of Copenhagen}%

\date{\today}% It is always \today, today,
             %  but any date may be explicitly specified

\begin{abstract}
Species interactions through cross-feeding via leakage and uptake of chemicals are important in microbial communities, and play an essential role in the coexistence of diverse species. Here, we study a simple dynamical model of a microbial community in which species interact by competing for the uptake of common metabolites that are leaked by other species. The model includes coupled dynamics of species populations and chemical concentrations in the medium, allowing for a variety of uptake and leakage networks among species.
Depending on the structure of these networks, the system exhibits different attractors, including fixed points, limit cycles, low-dimensional chaos, and high-dimensional chaos. In the fixed-point and limit-cycle cases, the number of coexisting species is bounded by the number of exchangeable chemicals, consistent with the well-known competitive exclusion principle. In contrast, in the low-dimensional chaotic regime, the number of coexisting species exhibits noticeable but limited excess over this limit. Remarkably, in the high-dimensional chaotic regime, a much larger number of species beyond this limit coexist persistently over time. In this case, the rank-abundance distribution
%shows a fat exponential tail,
is broader than exponential, as often observed in real ecosystems. The population dynamics displays intermittent switching among quasi-stationary states, while the chemical dynamics explore
%the high-dimensional space almost uniformly.
most of the high dimensions.
We find that such high-dimensional chaos is ubiquitous when the number of uptake chemicals is moderately larger than the number of leaked chemicals. Our results identify high-dimensional chaos with intermittent switching as a generic dynamical mechanism that stabilizes coexistence in interacting systems. We discuss its relevance to sustaining diverse microbial communities with leak-uptake cross-feeding.

\end{abstract}

%\keywords{Suggested keywords}%Use showkeys class option if keyword
                              %display desired
\maketitle

\section{Introduction}
Coexistence of diverse species under complex dynamics is a key topic in ecological systems. Traditionally, ecological dynamics have often been investigated by using population dynamics of species with direct species-species interactions. Microbial ecosystems that have gathered much attention recently, however, are not in such situation. The interaction among each species is mediated by chemical resources, and the combined dynamics between chemical concentrations and populations of species is required. How does such interplay between chemical and population dynamics increase the complexity in dynamics with diverse species?

In fact, competition for chemical resources among species has been analyzed by classic MacArthur model~\cite{MacArthur_AmNat1967}. Limit of the number of coexisting species competing for resources has been investigated both experimentally and theoretically~\cite{CuiMarslandMetha_PRL2020, MoranTikhonov_PhysRevX2022, BlumenthalMetha_PRL2024}.
In such systems, there is an established relationship between the number of possible coexisting species and the number of resources, as known as the competitive exclusion principle or also as Gause's law, which states that the former is bounded by the latter~\cite{Gause1932, Hardin1960, MacArthur_AmNat1967}. Coexistence of diverse species under limited resources, which is sometimes observed, therefore, has been discussed as ``paradox of the plankton''~\cite{Hutchinson_AmNat1961}.

In addition to resource competitions, microbes often leak chemicals which can be taken up and used by others. The interaction through the uptake and leakage of metabolites (cross-feeding) has recently gained increasing attention. 
In microbial communities, diverse species or strains coexist while secreting and exchanging hundreds of  metabolites~\cite{DattaCordero_NatComm2016, ZelezniakPatil_PNAS2015}. Even when the supply of resources is limited to single or few chemical components, exchange of chemical secreted by microbes can create new niches with each other, and thereby, diverse species can coexist~\cite{GoyalMasrov_PRL2018, GoldfordTikhonovMehtaSanchez_Science2018, DSouzaKost_NatProdRep2018, RosenzweigAdams_Genetics1994}.
Such exchange of chemicals can shape mutually symbiotic relationships, rather than parasitism among the microbial species~\cite{PonomarovaPatil_CellSystems2017, HilleslandStahl_PNAS2010, WintermuteSilver_MolSystBiol2010}.

For instance, a possible merit of cross-feeding has been theoretically investigated in a system with a detailed cell dynamics model~\cite{YamagishiSaitoKaneko2021}. The model includes intracellular dynamics, which is less coarse-grained but is numerically costly for studying the effect of cross-feeding on larger systems. 
Generally, the diversity of species can increase with the number of exchanged chemical species, as studied theoretically~\cite{GoyalMasrov_PRL2018}. Importantly, a network-based approach to the bipartite topology of cross-feeding has shown the emergence of bistability in diversity~\cite{Clegg-Gross_PNAS2025}. However, this further simplified framework so far has taken neither the competition for the leaked metabolites nor dynamical states into account. It has not yet been explored how competition for the leaked chemicals may lead to complex chemical and population dynamics and how such dynamics can shed novel light on the competitive exclusion principle.
In the cross-feeding systems, leaked metabolites can be resources of other species, and thus the bound on the coexisting species will generally include the number of exchangeable chemicals by species. The question to be addressed, then, is whether the complex dynamics may further enhance the number of coexisting species.

Here we introduce a minimal model of population dynamics to discuss communities with cross-feeding interactions with chemical dynamics. With this model, we demonstrate that cross-feeding can lead to complex dynamics both in population of species and chemical compositions, which may allow for successive alternating species. We will then show that such complex dynamics may allow for diversification of coexisting species, beyond the total number of exchangeable chemical components.

\section{Results}
\subsection{Dynamical leak-uptake model}
To model systems with both leakage and uptake of exchangeable chemicals, we start from the standard MacAuthor model~\cite{MoranTikhonov_PhysRevX2022} in which $N$ microbial species are competitively using $C$ chemical species (substrates, metabolites, etc.):
\begin{equation}
\dot{x}_\mu = \left( - \chi_\mu + \sum_{i=1}^{C}  \frac{\tau_{\mu i} b_i}{1 + T_i/K_i} \right) x_\mu,
\quad T_i = \sum_{\nu=1}^{N} x_\nu \tau_{\nu i}
%	\dot{x}_\mu = \left( - \chi_\mu + \sum_{i=1}^{C} \tau_{\mu i} h_i \right) x_\mu,
\label{Eq_MacAuthorCompetitionModel}
\end{equation}
where $\tau_{\mu i}$ is the matrix that represents the capacity of species $\mu$ to uptake chemical $i$, and $\chi_\mu = \chi_0 + \sum_i \chi_i \tau_{\mu i}$ is the base living cost of species $\mu$ which includes the cost of having the uptake capacities. Through the competition, the benefit of the chemical $i$ is reduced from its ideal benefit $b_i$ as the total uptake of that chemical over its carrying capacity $K_i$ increases.

%%%
\begin{figure}[bthp]
\includegraphics[width=1.0\linewidth]{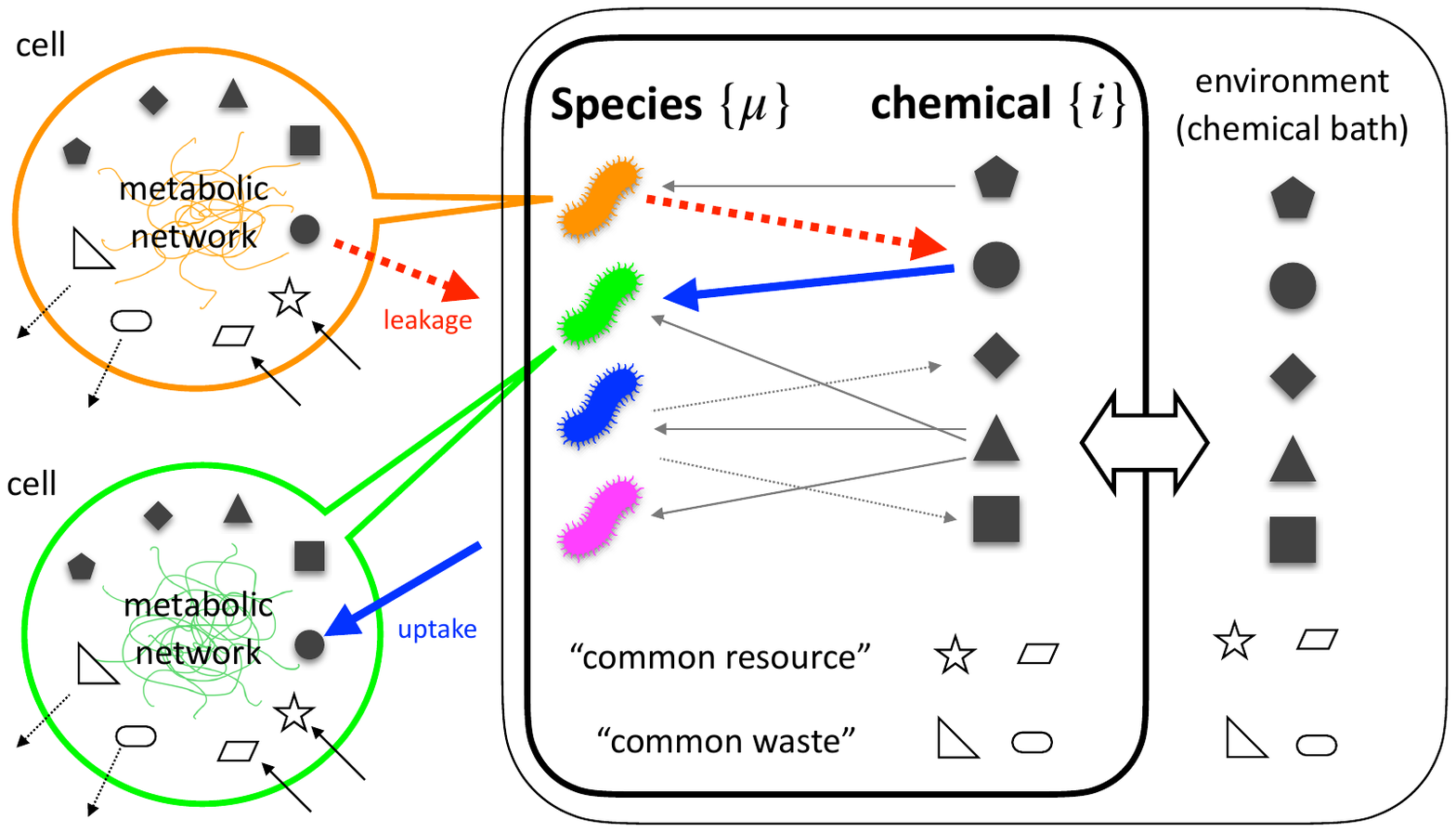}
\caption{
The schematic of the model ecosystem interacting through leak and uptake of chemicals, defined by Eqs.~[\ref{TSKKmodel_minimal}]. The filled symbols represent the chemicals those are leaked and taken by at least one of the species, and the open symbols are for the chemicals those are involved in the metabolic network inside the cells but only taken or leaked by them (i.e. common resource and waste, respectively).
}
\label{fig_model}
\end{figure}
%%%

When cells cross-feed through the leak and uptake of $C$ exchangeable chemical components which are a subset of the entire set of chemicals needed in their metabolic processes (Fig.~\ref{fig_model}), we need to consider the dynamics of the concentrations of chemicals.
Assuming that both the leakage and uptake of the chemicals by the cells are passive (diffusive) and the relaxation dynamics of chemical concentrations is much faster than the growth rates of each population, we can obtain a population dynamics model similar to Eq.~\ref{Eq_MacAuthorCompetitionModel}, with modified benefit function (see Appendix A1 for the details of the derivation).
For simplicity, we assume that cells gain no benefit from leakage itself and that the exchange of chemicals with environment is slower than the cross-feeding in the system. Then the population dynamics model is given as
\begin{eqnarray}
	\dot{x}_\mu
    &=&
    \left( - \chi_\mu + \sum_i^C \frac{\tau_{\mu i} b_i \bar{c}_i}{1 + U_i/K_i} \right) x_\mu,
    \cr
    \bar{c}_i
	&=&
    \frac{L_i}{U_i}
    = \frac{\sum_\upsilon^N x_\upsilon \lambda_{\upsilon i}}{\sum_\upsilon^N x_\upsilon \lambda_{\upsilon i} + \sum_\nu^N x_{\nu} \tau_{\nu i}}.
    \label{TSKKmodel_minimal}
\end{eqnarray}
where $\bar{c}_i$ is the equilibrium concentration of the chemical $i$ which is determined by the cell populations at that time. $\lambda_{\xi k} $ denotes the  leakage rate of chemical $k$ by the species $\xi$, and $L_i$ and $U_i = T_i + L_i$ are the amount of total leakage and total exchange of chemical $i$. The saturation factor with respect to $U_i$ is resulted from the natural assumption that the increase in exchange should be accompanied by the increase of the need for leak and uptake of chemicals, which are not explicitly treated in the leak-uptake relation but are conjugate with other chemicals through the metabolic process of each cells (chemicals represented by white symbols in Fig.1).

In addition to this population dynamics, we also introduce a lower limit $\epsilon = 10^{-16}$ to each population to include the base-level population flow by tiny migration and/or mutation~\cite{MallminMonte_PNAS2024, BlumenthalMetha_PRL2024}. By this treatment, no species will really go extinct but we treat only the species those population are greater than a detection limit as present.

\subsection{Typical behavior of the leak-uptake model}
\begin{figure}[htbp]
\includegraphics[width=1.0\linewidth]{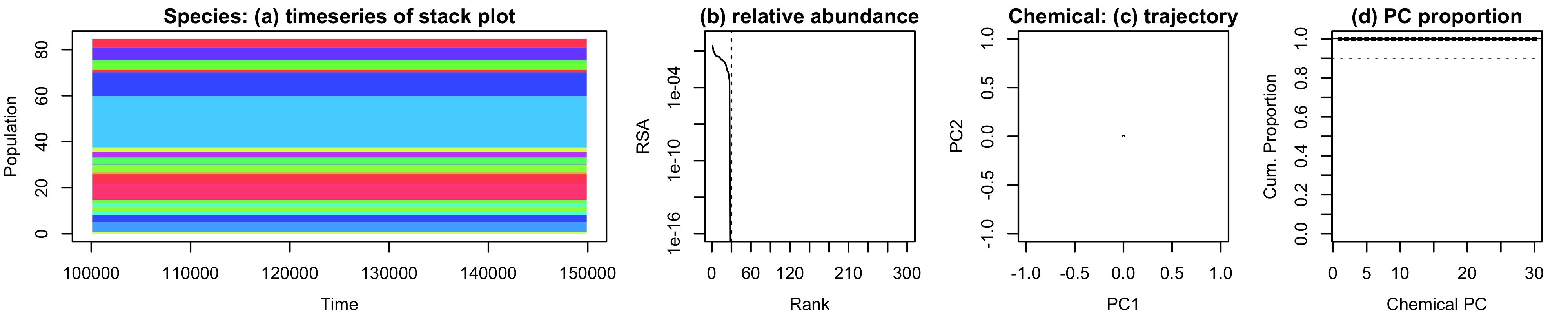}
\includegraphics[width=1.0\linewidth]{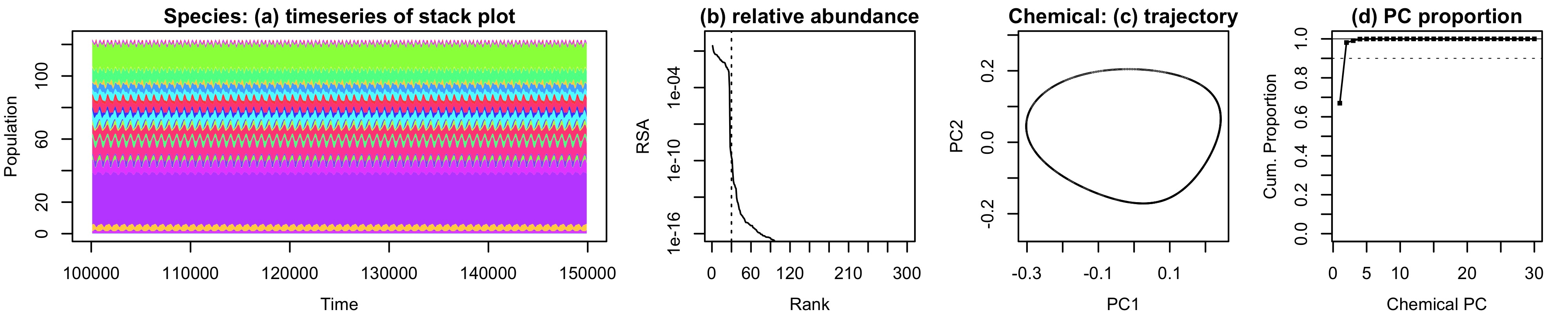}
\includegraphics[width=1.0\linewidth]{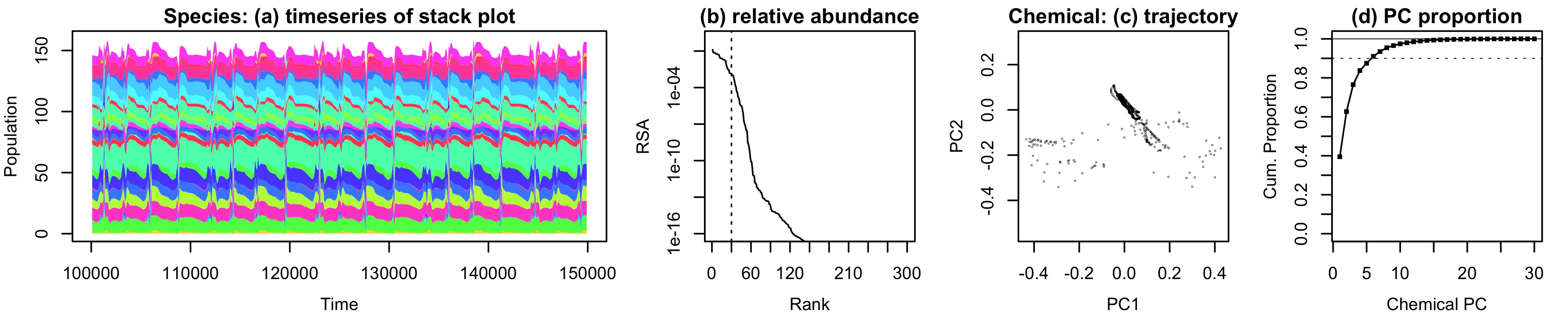}
\includegraphics[width=1.0\linewidth]{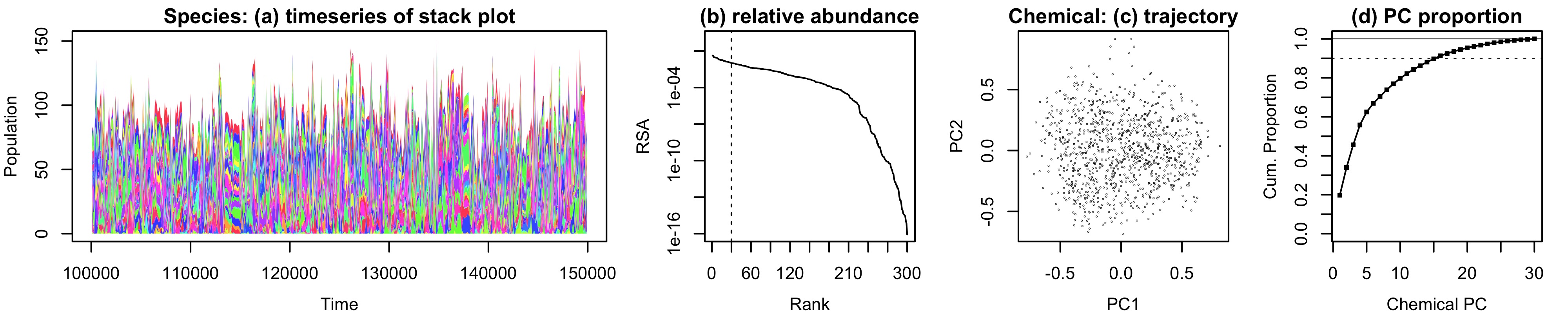}
\caption{
Typical behaviors of the system classified as fixed point, limit cycle, low-dimensional chaos, and high-dimensional chaos, respectively from top to bottom.
Panels in each row show the following information of each sample:
(a) Time series of the species population in a stacked plot,
(b) Relative species abundance based on the cumulative populations observed over the sampling period, the species,
(c) Trajectory of chemical dynamics in the plane defined by its first and second principal components, and
(d) Cumulative explained-variance ratio plot from PCA of chemical dynamics.}
\label{fig_behaviors}
\end{figure}

We have carried out simulations of Eq.~\ref{TSKKmodel_minimal}, taking randomly generated networks for a given number of chemical uptakes $m_{\rm take}$ and leaks $m_{\rm leak}$. The total number of species $N$ and that of chemicals $C$ are chosen to be $300$ and $30$ below, whereas the behaviors by changing $N$ and $C$ are explained later (also see Appendix B5 and B6). 
$N \gg C$ is assumed so that potential number of species is set to be much larger than the number of exchangeable chemicals.
Here we first set the parameters for basic cost of species uniformly at $\chi_0 = 0.1$ and $\chi_i = 0.5$ following~\cite{MoranTikhonov_PhysRevX2022} and for simplicity, and also set the carrying capacity of chemicals uniformly at $K_i = 1$. We assign the benefit of the chemical randomly as $b_i \sim b_0 \left| {\cal N}(0, 1) \right|$, where ${\cal N}(0, \sigma^2)$ denotes the normal distribution with average $0$ and variance $\sigma^2$. The scale of the benefit $b_0 = 100$ is chosen so that it is well larger compared to the basic cost $\chi_i$.
For each species $\mu$,  $\tau_{\mu i}$ for $m_{\rm take}$ components and $\lambda_{\mu i}$ for $m_{\rm leak}$ components are  assigned non-zero values, 
which are drawn randomly from distribution $\left| {\cal N} (0, 1) \right|$. Other matrix elements of $\tau$ and $\lambda$ are set to be $0$.

For a given randomly generated network, we initially set a small and random population $x_{\mu} > 0$ for all species. After some transients, only a part of those species exist above the detection limit $\delta$ \footnote{The detection limit $\delta$ is set at a certain level which is much smaller than the populations of major species and is much higher than the imposed lower limit, typically at $\delta = \sqrt{\epsilon}$. The result in the following is not essentially changed as long as $\epsilon$ is enough small and the $\delta$ satisfy the same condition, for example, $\epsilon \in [10^{-12}, 10^{-20}], \ \delta = \sqrt{\epsilon}$.}.
Here we discuss the behavior of population and chemical dynamics, after initial transients $t \in (0, 10^5)$ are discarded.

In Fig.~\ref{fig_behaviors}, we display typical examples of the population and chemical dynamics. In (a), the timeseries of population of species is plotted whereas the average abundances of species are shown as rank-abundances plot in (b). The dynamics of chemical concentrations are plotted as the trajectory in the chemical phase space by using the first and second principal components (in (c)), whereas accumulated contribution over the principal components are shown in (d). 
As shown in the plots, the behaviors of population/chemical dynamics are classified as follows:
\begin{itemize}
    \item[(i)]
    Fixed point: Both the population and chemical dynamics fall onto fixed point(s). In some cases, a unique fixed-point attractor is reached from different initial conditions (as far as we checked), and in other cases there are multiple fixed-point attractors (see Fig.~\ref{SI_fig_multiFP} in Appendix B). The number of coexisting species is bounded by the number of chemical components $C$ ($=30$).
    
    \item[(ii)]
    Limit cycle: The population and chemical concentrations exhibit a cyclic oscillation, as can be seen in the population time-series and the trajectory in chemical space (in some cases, multiple limit-cycle attractors and fixed-point attractors coexist). As in the fixed-point case, the number of remaining species does not significantly exceed $C$, as is consistent with the exclusion principle.
    
    \item[(iii)]
    Low-dimensional chaos: Population and chemical dynamics exhibit chaos. The trajectory of chemical composition shows a low-dimensional orbit and falls on an attractor within a few (typically $3 \sim 7$) dimensions, and thus the accumulated contribution of principal components goes close to unity up to $7$,
    which is much fewer than $C$.
    The number of species is bounded by $C$ for some time, but can intermittently go beyond $C$, and thus can break the Gause's limit transiently,
    but this excess is limited within a restricted range compared with the whole prepared species.

    \item[(iv)]
    High-dimensional chaos: Species grow and go ``extinct'' alternately, leading to the complex population dynamics across all species. Here the number of species that appear in the timeseries goes much beyond $C$, as shown in the rank-abundance plot and is extended to almost fully to the prepared species. The dynamic change of chemical components covers almost the full dimension of $C$, as shown by the accumulated contribution in the principal components. In the state space of chemical concentrations, no salient structure is observed: as shown in the bottom panel (c) of Fig.~\ref{fig_behaviors}, the points are uniformly scattered within a high-dimensional cuboid.
\end{itemize}

We estimate the dimension of chemical dynamics $d_c$, by using the principal component analysis and computing how many components are required for the cumulative contribution of, say, $90\%$.
The fixed point case is trivially $0$~dimension, whereas the dimension in the case (ii) is typically less than $3$. The distinction between (iii) and (iv) is also associated with $d_c$, as $d_c = 3 \sim 7$ in the former case and is typically beyond $7$ in the latter. In some cases, a network has multiple attractors of different dimensions (see Appendix B1).

Although we displayed one example for each class in Fig.~\ref{fig_behaviors}, the validity of the above classification is confirmed by extensive simulation across a variety of networks, as partly shown in Appendix B2, B3, and B4.

\begin{figure}[htbp]
\includegraphics[width=1.0\linewidth]{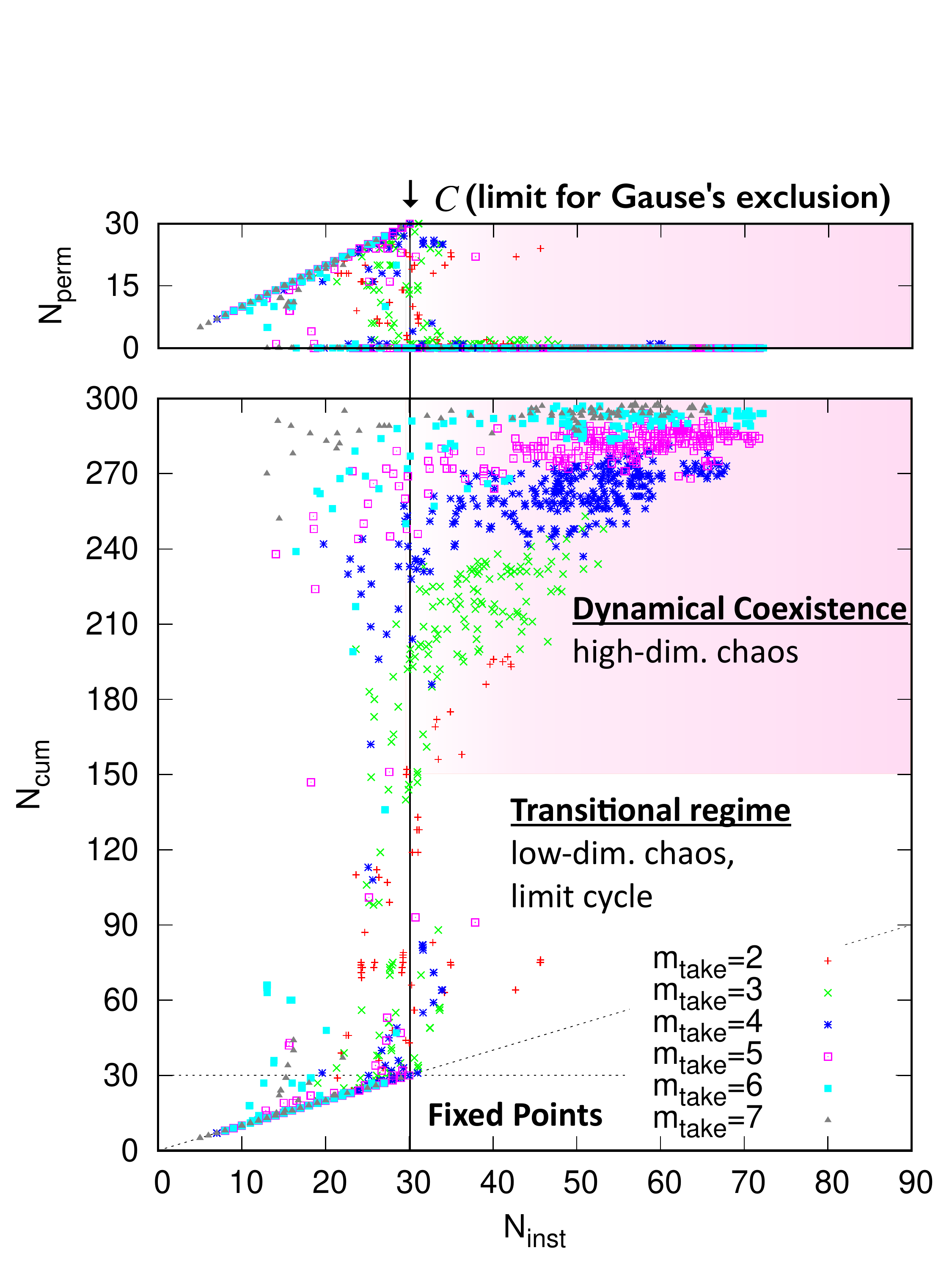}
\caption{
Transitional change in $N_{\rm inst}$ (the temporal average of instant number of coexisting species at each moment) and $N_{\rm cum}$ (cumulative number of species which appears in the observation time) to the change in $m_{\rm take}$ (the lower main panel).
In the upper sub panel, the corresponding relation between $N_{\rm inst}$ and $N_{\rm perm}$ (number of species those are stably observed in the whole observation time) is shown. System parameters are fixed at $N = 300, C = 30, m_{\rm leak} = 2$. In the dynamically coexisting regime ($N_{\rm cum} \gg C \cap N_{\rm inst} > C$), only a small minority of species show stable presence ($N_{\rm perm} \approx 0$) indicating that the set of species above the observation threshold keep alternating.
}
\label{fig_Ninst-Ncum}
\end{figure}
%Transitional change in $N_{\rm instant}$ (number of coexisting species at each moment) and $N_{\rm total}$ (total number of species which appears in the observation time) to the change in $m_{\rm take}$ (center panel), and the corresponding typical dynamics of each regime (surrounding panels). Trajectories are shown in the PCA plane of chemical space, together with the cumulative proportion of PCs (PC prop.) and the relative species abundance of the population (RSA). System parameters are fixed at $N = 300, C = 30, m_{\rm leak} = 2$.
\begin{figure}[htbp]
\includegraphics[width=1.0\linewidth]{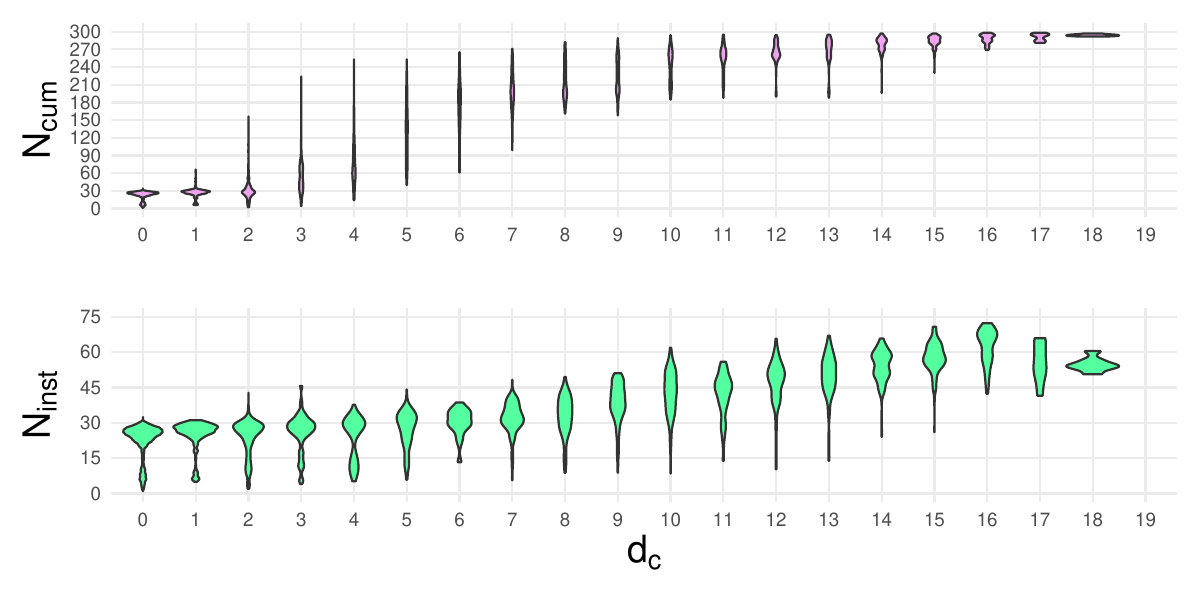}
\caption{
Distributions of cumulative number of species ($N_{\rm cum}$, top) and instant number of species ($N_{\rm inst}$, bottom), to the effective dimension of the chemical dynamics of corresponding timeseries, $d_c$. While $N_{\rm cum}$ starts to exceed the Gause's limit $C=30$ at around $d_c=3$, $d_c = 6$ or higher dimension is needed to overcome the Gause's limit in typical $N_{\rm inst}$. 
}
\label{fig_NtoDim}
\end{figure}

\subsection{Exceeding the competitive exclusion principle}
Next, we investigate how the number of coexisting species changes in time, in more detail. As already mentioned in the introduction, the number of species that coexist stably is limited by the number of resources, as known as Gause's exclusion principle. In the present model, the resources are not supplied explicitly from the surrounding environment but are created by species. Hence the number of exchangeable chemical species ($C=30$ in the above example) will limit the number of species that stably coexist. In fact, this is valid as long as the state is stationary: In the case of fixed points or a limit cycle, the number is limited by the number of chemical species $C$.
% ちょっと英語変かもですがそしてそこで書くか迷いましたが、1-A の末尾に書いてしまったのでここはコメントアウトしました。検出限界の設定はダイナミクスの本質ではないのでまだ迷ってますが(TS)
%~\footnote{Here we assume that the species exist when its population fraction $x_{\mu}$ is larger than $\epsilon = 10^{-8}$ Still the results below do not change essentially for this range of threshold $\epsilon = [10^{-?}, 10^{-?}]$.}

In the chaotic states (iii) and (iv), however, the number of species can significantly exceed $C$, whereas the existing species can alternate in time. To examine the dynamics of existing species,
%then,
%\footnote{Here, we examine which species exist (beyond the population threshold..), after discarding the initial transient.}
we need to distinguish the following three quantities;

\begin{itemize}
\item 
The number of species that coexist at each instant, $N(t)$. As this can vary in time, we take the average over time, $N_{\rm inst} = \overline{N(t)}$.
\item 
The number of species that appear at least once in the whole observation time (the number of the cumulative set of distinct species), $N_{\rm cum}$.
\item
The number of species that exist permanently throughout the observation time, $N_{\rm perm}$.
\end{itemize}

From the definition, $N_{\rm perm} \leq N_{\rm inst} \leq N_{\rm cum}$ holds.
In Fig.~\ref{fig_Ninst-Ncum}, we have plotted $N_{\rm cum}$ and $N_{\rm perm}$ versus $N_{\rm inst}$ across 50 different networks for different values of $m_{\rm take}$ and $m_{\rm leak}$, as well as for different initial populations. First, as expected, for the cases of low-dimensional attractors (case (i) and (ii)), $N_{\rm cum} = N_{\rm inst} \leq C$ is satisfied, as shown in the figure. This follows the Gause's limit. 

Now, for more complex dynamic attractors, (iii) and (iv),
both $N_{\rm inst}$ and $N_{\rm cum}$ can go beyond $C$, where two regimes are observed, as seen in the bottom panel of Fig.~\ref{fig_Ninst-Ncum}.  In the case (iii), that typically corresponds to the ``transitional regime'', where $N_{\rm inst}$ is not so large (say up to $\sim C+10$), $N_{\rm cum}$ is extended to much larger than $N_{\rm inst}$ which is still much lower than the total number of prepared species $N$. Hence, the species alternate over time %generations,
but are not extended to all possible species.
Note that in this case, $0 < N_{\rm perm} < C$ holds. %Even though
While the number of instant or cumulative species goes beyond the Gause's limit, the species that permanently exist over time is much less than $C$.

%As $N_{\rm cum}$ further increases, $N_{\rm perm}$ further decreases.
%even in this regime it is above 0, implies that some species permanently exist over time.

For the high-dimensional chaos (iv), $N_{\rm inst}$ is much larger, and $N_{\rm cum}$ approaches to the total $N$. In fact, in this case, if we increase $N$, $N_{\rm cum}$ increases by keeping the $N_{\rm cum} \sim N$, where $N_{\rm inst}$ also increases with it (see Appendix B5).
In contrast, $N_{\rm perm}$ often goes to zero, and is few at maximum (Fig.~\ref{fig_Ninst-Ncum} top).
To sum up, for the high-dimensional chaos case, the number of species that exist at each instant time is far beyond the Gause's limit $C$, whereas they alternate in time
%to penetrate to the whole range of
covering most of the species prepared. The existing species change over time, and there are typically no species that permanently exist over time.

%Note that even if the number of species is constantly beyond the Gause's limit in the chaotic case, each species does not persist. and $N_{always}$ is lower than $M$.  Rather it decreases as $N_{inst}$ and $N_{cum}$ increase.$  In the high-dimensional chaos with $N_{cum}\sim N$, $N_{[always}$ turns to be zero. No persistent species exist and the existeince species itinerate over the whole speiceis $N$ that are preparedto.

Now, to elucidate how the emergence of each phase (i)-(iv) depends on the number of co-existing species above, we examined the relationship between $N_{\rm cum}$ or $N_{\rm inst}$ with the chemical dimension $d_c$.  
In Fig.~\ref{fig_NtoDim}, we plot $N_{\rm inst}$ and $N_{\rm cum}$ against the chemical dimension, across different trajectories (i.e. different networks and initial conditions). As shown, $N_{\rm inst}$ and $N_{\rm cum}$ are equal to or less than $C$, when chemical dimension is smaller than 3, whereas they increase monotonically with the chemical dimension. When the dimension is larger than 7, that corresponds to the high-dimensional chaos, $N_{\rm cum}$ approaches $N$ and $N_{\rm inst}$ is typically larger than $C$ significantly. These validate the above discussion and the correspondence of classification by the chemical dynamics with the coexisting population dynamics.

\subsection{Rank-abundances plot in the high-dimensional chaos case}
\begin{figure}[bthp]
\includegraphics[width=1.0\linewidth]{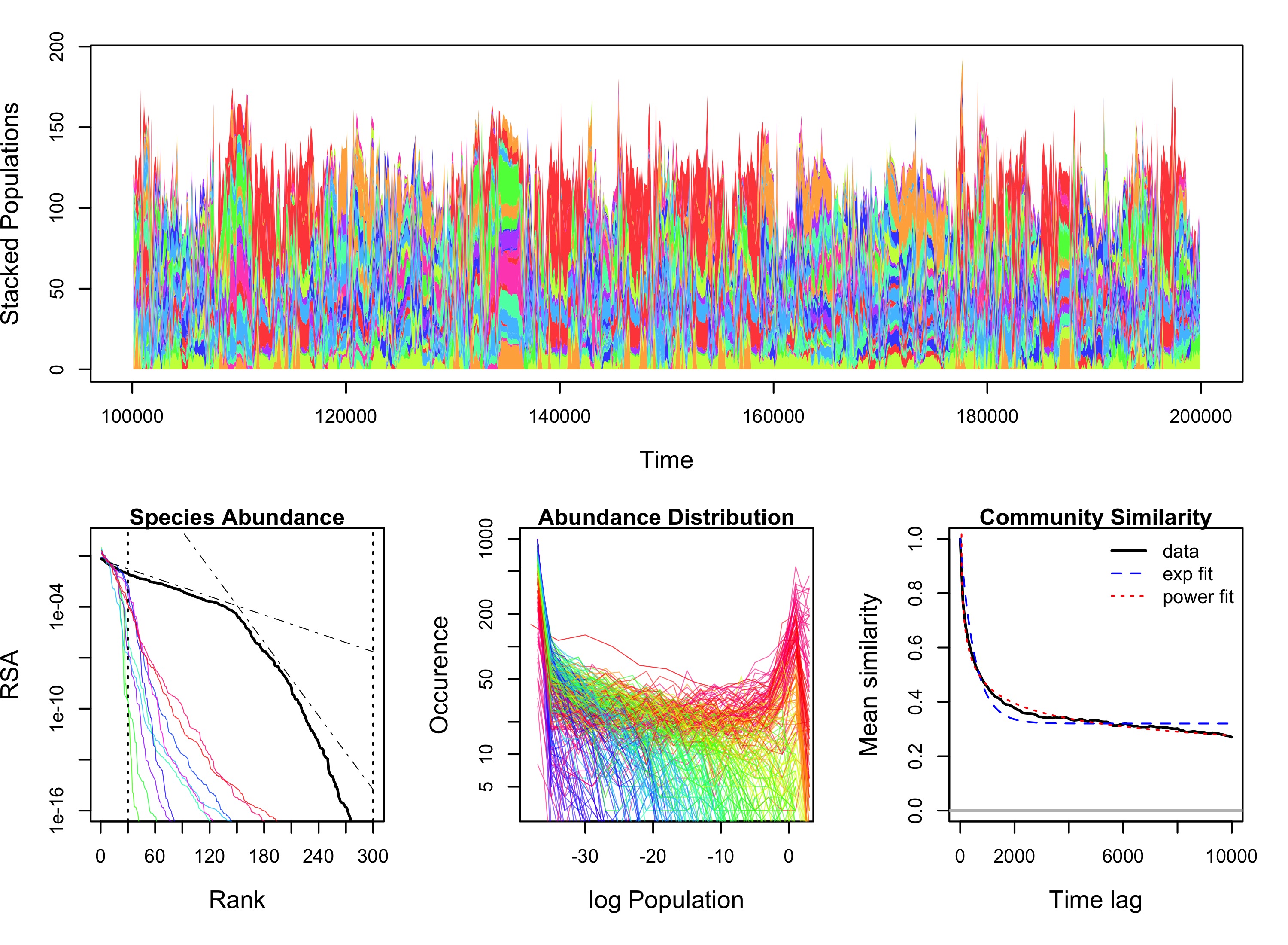}
\caption{
A typical example of the dynamics classified as (iv) high-dimensional chaos.
(Top) The timeseries of stacked species populations, which shows an intermittent dynamics among different community structures.
(Bottom left) Relative species abundance (RSA) curves of cumulative populations (bold black line) and instantaneous populations at different moments (colored thin lines). In this regime, instantaneous RSA also becomes broader beyond Gause's limit ($C=30$, shown by the vertical dotted line).
(Bottom center) Distribution of the instantaneous populations of species. Colors from red through yellow and green to blue represents the rank of that species in the cumulative population. It shows that top major species often take large population size but also sometimes take very small population, and the species with following majority in average take large population less frequently. 
(Bottom right) Temporal decay of the community similarity. In consistent with the intermittent feature of the population timeseries, community similarity shows a power-law decay in time as $\sim t^{-0.25}$.
}
\label{fig_RSAandSimilarity}
\end{figure}

As many species emerge and disappear through time in the high-dimensional chaos, it is suggested that there are no fittest species over the whole range of time. The next question to be addressed, then, is the population size distribution. In Fig~.\ref{fig_RSAandSimilarity}, the rank-abundance relationship is plotted (bottom left panel)

The rank-abundance plot at each instance follows an approximately the exponential form with a %thicker
broader tail, as shown by the colored thin lines in Fig.~\ref{fig_RSAandSimilarity}.
The exponential rank-abundance relationship is consistent with the geometric sequence model~\cite{Motomura1932, DoiMori_Oikos2013}, whereas the existence of a
%thicker tail can be
broader tail is consistent with the distribution expected from other theories and observations~\cite{Preston1962, VolkovHubbellMaritan2003, Shoemaker2017, GoyalMasrov_PRL2018}.
Note that each species has different uptake and leakage networks, and its intrinsic growth rate is not equal. In this quasi-stationary state, however, the growth rate is approximately balanced, leading to a behavior closer to neutrality. However, this balance is not complete and the population of each species exhibits alternations of exponential growth and extinction as in the top panel of Fig.~\ref{fig_RSAandSimilarity}.

The alternation of abundant species then leads to the change in the rank-abundance plot; higher rank species may become invisible and are replaced by emerging species. To examine the long-term behavior, we have computed the cumulative abundance over a long time span, and plotted against their rank, also in Fig.~5 (bold blak line in the bottom left panel).
As shown, the plot shows double exponential decay: ${\rm exp} \{ -C_1 r \}$ for low rank  $r$ and ${\rm exp} \{-C_2 r \}$ for higher rank. with $C_1 \approx 1/24 < C_2 \approx 1/6$ as constants. A possible origin of this is the existence of two types of species with long and short presence time. In the bottom center of Fig.~5, the instant population distribution of each species is plotted, by changing colors according to the ranking in temporal average. As shown, the higher rank-species exhibit a double peak: They keep high abundances for a sufficient time span, whereas they also go minor below the observation threshold for some time span.
In fact, even in the high-dimensional chaos, quasi-stationary community sometimes appears as shown in Fig.~\ref{fig_RSAandSimilarity} (top).
Here, those species that participate in the frequently appearing community structures and that participate in the long-term quasi-stationary regime maintain presence as dominant species for a long time span.

The existence of long-term quasi-stationary states and their alterations suggest the punctuated-equilibrium behavior predicted by the self-organized criticality models~\cite{BakSneppen1993, Flyvbjerg1993}, the tangled-nature models~\cite{Christensen2002, MuraseRikvold_JTB2010}, and the chaotic itineracy~\cite{ChaoticItineracy_Chaos2003}.
To examine the long-term correlated behavior, we have computed how the similarity in population distribution decays in time (see Materials and Method for definition). As shown in the bottom right of Fig.~\ref{fig_RSAandSimilarity}, we observe power-law decay in time, as $\sim t^{-0.25}$.
In contrast, for low-dimensional attractors, the rank-abundance relationship deviates from such broad-tailed behavior, as has been shown in Fig.~\ref{fig_behaviors}.
The population size  highly depends on each species. 

\subsection{Dependence of frequency of the four phases upon the number of takes and leaks}

\begin{figure}[tbhp]
\includegraphics[width=1.1\linewidth]{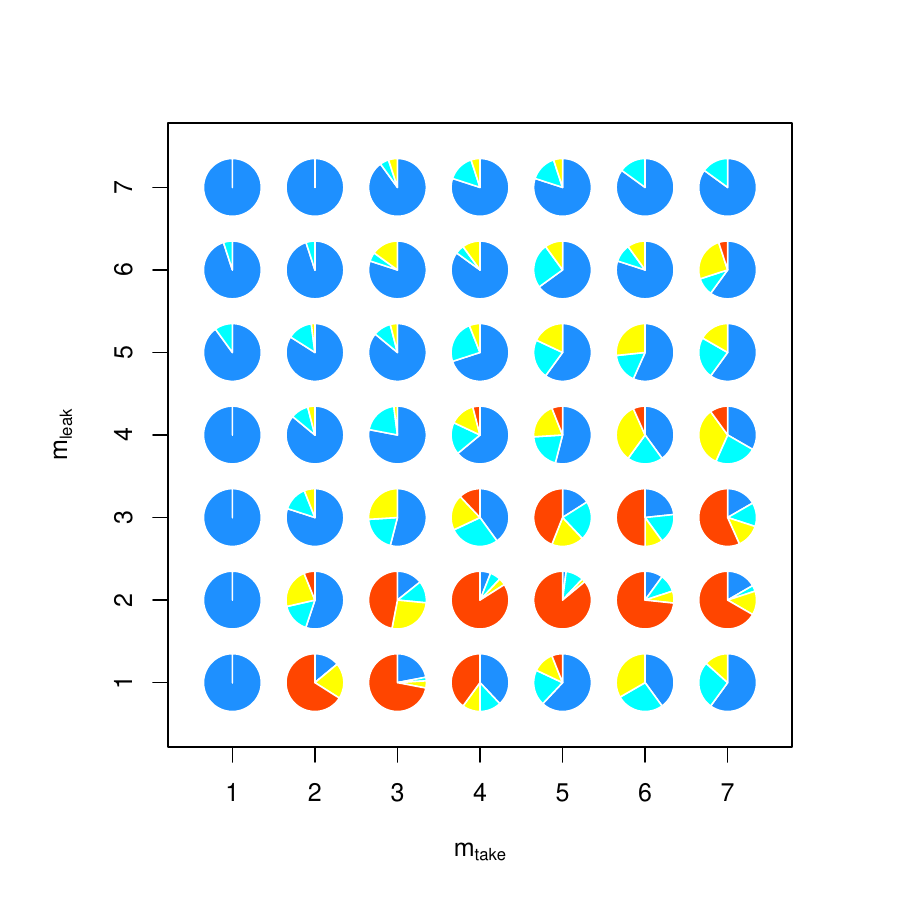}
\caption{
Phase diagram of the leak-uptake model in the $m_{\rm leak}-m_{\rm take}$ plane. Blue, cyan, yellow and red segments represents the proportion of the networks whose most complex dynamics out of 10 samples is classes (i), (ii), (iii), and (iv), respectively.
System size is fixed at $N = 300, C = 30$ and $50$ network samples are taken for each parameter set.
}
\label{fig_PhaseDiagram}
\end{figure}

Which of the cases (i)-(iv) emerges depends on the structure of specific leak-uptake network, where the number of leaks and the number of uptakes is a dominant factor. We thus study how the statistical frequency of the cases (i)-(iv) depends on the number of uptake $m_{\rm take}$ and leak $m_{\rm leak}$ of chemicals per species, by taking $50$
randomly chosen networks for given  $m_{\rm take}$ and leak $m_{\rm leak}$. 
Following the results in section C, each of the four classes is categorized by the  maximal chemical dimension $d_c$ among its trajectories starting from ten different initial conditions
(see Appendix A2 for the histogram of $d_c$ itself before classification).
As shown in Fig.~\ref{fig_PhaseDiagram}, as the number of uptakes $m_{\rm take}$ increases, the fraction of networks that exhibit complex dynamics (in particular, high-dimensional chaos) is more frequent. An appropriate fraction of $m_{\rm take}/m_{\rm leak}$ seems to be needed (i.e., larger than $1$ but not too large), to have high-dimensional chaotic behavior. As the number of leaks increases $m_{\rm leak}$, the fraction of networks with fixed points increases. For the regime of $m_{\rm leak} > m_{\rm take}$, no chaotic behaviors are observed
(see Appendix B6 for the diagram for larger $N$ and B7 for the one after longer transient time.)
To sum up, competition on leaked resources is the dominant factor for complex dynamics.

\section{Discussion}
%\begin{enumerate}
 %   \item Violation of the principle of competitive exclusion by highly chaotic dynamics.
  %  \item Resource competition systems can have chaotic dynamics~\cite{BlumenthalMetha_PRL2024}.
   % \item Emergent neutrality and broad relative abundance of species~\cite{MallminMonte_PNAS2024}.
%\end{enumerate}

In the present paper, we have numerically investigated an ecosystem model with leak-uptake exchange of chemicals.
When the numbers of uptake and leak chemicals per species are moderate and asymmetric ($m_{\rm take} > m_{\rm leak}$), the mutual exchange of chemicals among species often leads to complex dynamics, where diverse species coexist in each time span, whose number is far beyond Gause's exclusion limit. 
The dynamics in this high-dimensional chaos look almost random at a glance:  
no structure in chemical-state dynamics is observed, and populations of all possible species appear and disappear in time, covering all the species prepared in the system.
Still, there are time spans in which the population of species is quasi-stationary, where cross-feeding among species support such quasi-stable states. After some time span, however, alternation of species follows, leading to rather chaotic dynamics with appearance and disappearance of novel species, until the next quasi-stationary regime with different species emerges.
Organized quasi-stationary behavior exists locally in time, but as a long-term, many species appear and disappear rather randomly.
Such high-dimensional chaotic behaviour is consistent with empirically observed chaotic behavior in plankton and other ecosystems such as insects~\cite{Srephen_PlanktonChaos_Nature2008,Sthephan_UbiquitusChaos_NatEcoEvo2022}.

It is interesting that leak-uptake dynamics that lead to high-dimensional chaotic dynamics support the coexistence of diverse species that go beyond Gause's limit, even for each instance in time.
It is noted that in the present cross-feeding leak-uptake system, there is no explicit external supply of resources and can be regarded as a limit of resource scarcity. Hence, under a limited external resource, stable ecosystem may be lost, but a resultant complex dynamics will support the diversity among species.
In such case, rank-abundance plot at each instance shows a fat tail beyond the exponential form. Many species keep the potential with partially neutral manner. Here, coexisting species themselves, including the high-rank species, will alternate in a long time span, but the rank-abundance distribution keeps the fat-tailed exponential form.
Sustainment of such broad rank-abundance distribution with the change in the rank itself is consistent with that observed for the soil microbiological ecosystem~\cite{Ecosoil_Shimada2024}. Existence of multiple quasi-stationary states in our model will also be relevant to the transitions observed therein.

% Rondom LV 系の chaotic turn over との関連について言及のパラグラフ
Previous studies have shown that strongly interacting Lotka-Volterra communities can display persistent non-equilibrium dynamics driven by rare species, In particular, Mallmin {\it et al.} reported chaotic turnover between rare and abundant species~\cite{MallminMonte_PNAS2024}, while Arnoulx de Pirey and Bunin described long-time fluctuations as an effective jump process associated with species hovering near extinction~\cite{PireyBunin_PRX2024}.
Here, we demonstrate that similar dynamics also arise in a natural resource competition model with cross-feeding, indicating that these phenomena are not specific to Lotka-Volterra formulations. Moreover, our results characterize how the system transitions from fixed point, through intermediate dynamical regimes, to a phase which is highly chaotic yet with temporal community structure and intermittent switching among them.

Alternate possibility to resolve the plankton paradox was proposed as the diversity introduced by the exchange of a huge variety of chemicals with energy constraint~\cite{GoyalMasrov_PRL2018}. In the present study, in contrast, even if the chemical diversity is not so huge, diverse species more than chemicals is supported by high-dimensional chaos.
Possibility of overcoming of the Gause's limit by dynamical state has been argued, by introduction of external noise, or oscillation or low-dimensional chaos (as in our case (ii) and (iii)) as a result of switching of species dynamics due to Liebig's laws of minimum~\cite{Huisman_DynamicalBreakingOfGause_Nature1999}, and so on. Here we show that asymmetric cross-feeding relation naturally leads to dynamical attractors which breaks the limit.
Chaotic turnover of coexisting species has previously been discussed by using generalized Lotka-Volterra equation. In particular, in~\cite{MallminMonte_PNAS2024}, resource-competition dynamics derived from MacArthur model~\cite{MacArthur_AmNat1967} can exhibit chaotic turnover, when the non-reciprocal interaction is dominant, under taking the limit of sufficient number of resource chemicals, that satisfies the Gause's condition. In contrast, in our study, both with the uptake and leak of chemicals are taken into account to support the cross-feeding where reciprocal interaction is dominant. Competition of leaked chemicals by species sharing the same uptake chemicals results in high-dimensional chaos, with the dynamic violation of Gause's exclusion principle.
Extension of dynamic mean-field analysis to cover the high-dimensional dynamics and population dynamics of diverse coexisting species will be an important step to be explored in future.

As for the cross-feeding by leak-uptake of chemicals, Clegg and Gross beautifully analyzed the fixed-point states and their stability, by generation function method. These states correspond to the stationary states (case (i)) in our model. Our study, in this sense, extends their result to the regime with dynamical attractors including high-dimensional chaos. An approach by dynamical network models~\cite{Takashi_EOS_SREP2014, Fumiko_EOS_SREP2017, Fumiko_EOS_RoySocOpenSci2019} may fill the gap.

Dynamics of microbial ecosystems has garnered much attention, which include biofilm, gut bacterial system, and  soil ecosystem. Often, diverse microbes coexist with the interaction by exchanging chemicals. By analyzing the long-term population dynamics together with the chemical dynamics, comparison with the high-dimensional chaotic state we uncovered here will be possible in future.

% 12/27 TS もういちど全部チェックしながらここまできた。

%%%%%%%%%%%%%%%%%%%%%%%%%%%%%%%%%%%%%%%%%%%%%%%%%%%%%%%%%%%%%%%%%%%%%%%%%%%
% Material & Methods
%%%%%%%%%%%%%%%%%%%%%%%%%%%%%%%%%%%%%%%%%%%%%%%%%%%%%%%%%%%%%%%%%%%%%%%%%%%
%Please describe your materials and methods here. This can be more than one paragraph, and may contain subsections and equations as required.

\begin{acknowledgements}
This work was supported by JSPS KAKENHI Grant Number JP23K03256 to TS and the Novo Nordisk Foundation Grant No. NNF21OC0065542 to KK. We thank Shigeto Otsuka, Namiko Miatrai, and Kim Sneppen for valuable comments on this project.
\end{acknowledgements}

%\clearpage
%\bibsplit[100]
%Use \bibsplit to split the references from the body of the text. Value "[2]" represents the number of reference in the left column (Note: Please avoid single column figures & tables on this page.)

\appendix

\section{Methods}
\renewcommand{\thesubsection}{\Alph{section}\arabic{subsection}}
\subsection{Derivation of the population dynamics model of cross-feeding systems}
Let us consider a system in which cells are cross-feeding through leak and uptake of $C$ exchangeable chemical species, which form a subset of the entire set of chemicals needed in their metabolic processes (Fig.~\ref{fig_model}).
To consider the benefit from the chemicals, we need to first treat the dynamics of the concentration of chemicals. Assuming that both the leakage and uptake of the chemicals are passive (i.e. diffusive), the dynamics of the chemical concentration in the common pool should be
\begin{equation}
	\dot{c}_i
	=
	- c_i \sum_{\nu}^{N} x_\nu \tau_{\nu i}  
	+ (1-c_i) \sum_{\upsilon}^{N}  x_{\upsilon} \lambda_{\upsilon i}
    + d_i (c^*_i - c_i),
\end{equation}
where $\tau_{\mu i} \ge 0$ and $\lambda_{\upsilon i} \ge 0$ are the species $i$'s capacity of uptake and leak of the chemical $i$, respectively. $c^*_i$ is the concentration of the chemical $i$ in the external reservoir surrounding the population cells and $d_i$ is the diffusion constant between the focal system and the reservoir.
Because the dynamics of chemical concentrations is generally much faster than the growth rate of each population, it is natural to assume that the chemical concentrations are at an equilibrium
\begin{equation}
	\bar{c}_i = \frac{L_i + d_i c^*_i}{T_i + L_i + d_i}
	\qquad
	\left( T_i = \sum_{\nu}^N x_{\nu} \tau_{\nu i}, \ L_i = \sum_{\upsilon}^N x_\upsilon \lambda_{\upsilon i} 
    \right)
\end{equation}
for a given population.
The rate of uptake of chemical $i$ for uptake should be proportional to the chemical concentration, and hence the ideal benefit is.

The benefit should also reach saturation as the total amount of the leak and uptake of the chemical increases. It is because the increase in the exchange of that chemical should be accompanied by the increase of the need for leak and uptake of other chemicals which are conjugate through the intra-cellular metabolic process, which are not in the leak and uptake relation (shown by white symbols in Fig.\ref{fig_model}). % which which になってしまったけれども
Therefore, even in a case that the leakage and uptake are nicely balanced and hence the chemical concentration is kept constant, larger amount of exchange should spoil the benefit of the cells. We model this effect by the carrying capacity $K_i$ for the total leak and take $U_i = L_i + T_i$ as
\begin{equation}
    h_i = \frac{b_i \bar{c}_i}{1 + U_i/K_i}.
\end{equation}
In the following, we assume that the exchange of the chemicals with the surrounding chemical bath is small compared to the speed of exchange in the focal system. We here also assume that the benefit of leaking a chemical is $0$ and the carrying capacity of the chemicals are uniform: $K_i=1$, for simplicity.
Then the population dynamics under this minimal setting is given as Eq. \ref{TSKKmodel_minimal}.

\subsection{Classification of the network dynamics}
\begin{figure}[htbp]
\includegraphics[width=1.0\linewidth]{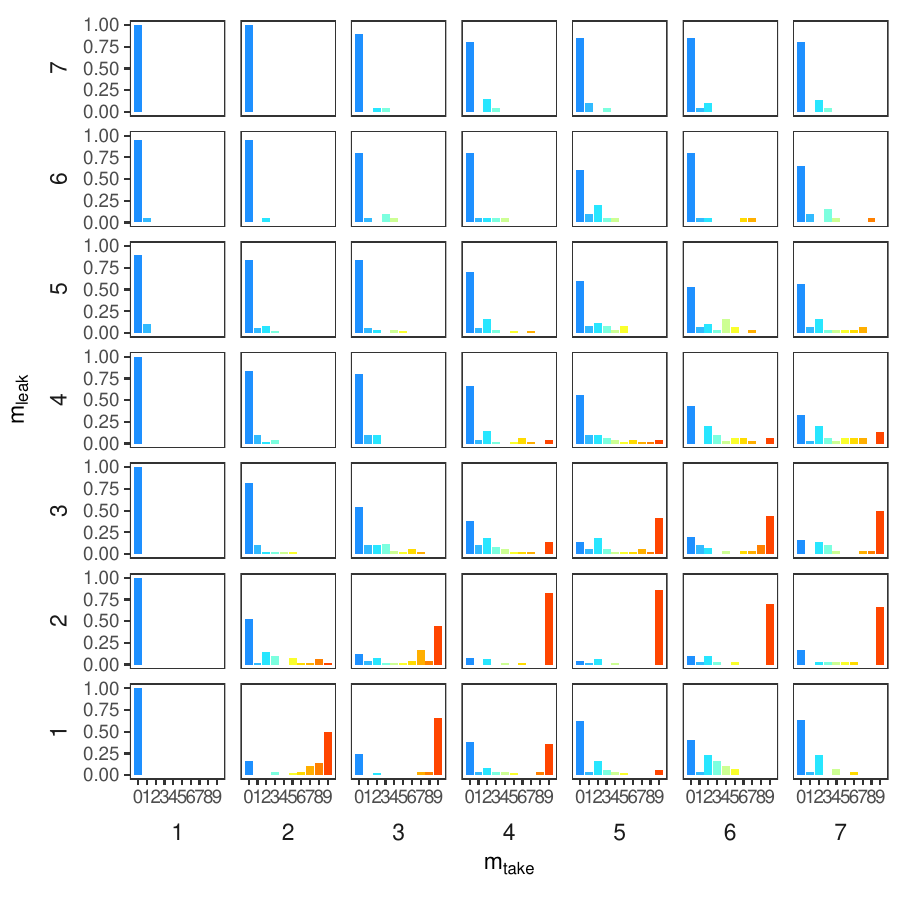}
\caption{Distributions of the maximum dynamical dimensions of the chemical trajectory $d_c^*$, obtained from the same data as Fig.~\ref{fig_PhaseDiagram}, i.e. system size is fixed at $N = 300, C = 30$ and $50$ network samples are taken for each panel.}
\label{fig_PDF_dc}
\end{figure}

For a given network parameter set $(N, C, m_{\rm take}, m_{\rm leak})$, we create multiple network instances. In generating the network, trivial self-feeding links ($\tau_{\mu i} \lambda_{\mu i} \neq 0$) and duplicate links are prohibited. For each network instance, we take samples from 10 different initial conditions to take account the possibility of having multiple attractors. We define the effective dimension of chemical dynamics by $d_c$ as the dimension at which the cumulative explained-variance ratio first exceeds $0.9$. Then each network instance is classified by the maximum effective dimension among the samples $d_c^* = \max\{ d_c \}_{\rm samples}$ as; $d_c^* = 0$: class (i) (fixed point), $0 < d_c^* \le 2$: class (ii) (limit cycle), $3 \le d_c^* \le 7$: class (iii) (limit cycle, low-dimensional chaos, or quasi-periodicity),
% 曲面上の limit cycle でも、今の主成分分析の数え方だと d_c > 2 になってしまうことがあるので
and  $8 \le d_c^*$: class (iv): (high-dimensional chaos). Note that, as shown in Fig.~\ref{fig_PDF_dc} the effective dimension around the boundary regime between (iii) and (iv) ($d_c \sim 7$) is sparse so that the change of criterion to 6 $\sim$ 9 does not alter our conclusion such as the phase diagram shown in Fig.\ref{fig_PhaseDiagram}.

\subsection{Relative Species Abundance}
Relative species abundance distributions are calculated for both instantaneous species population $\vec{x}(t) \equiv \left( x_1(t), \cdots x_\mu(t), \cdots x_N(t) \right)$ and cumulative population over the sampling time points $\vec{x} = \left( \left\{ \sum_{t \in T_{\rm obs.}} x_\mu(t) \right\} \right)$. 

\subsection{Community Similarity and its Decay}
Similarity between the communities at different moments is measured by the product of normalized populations as,
\begin{equation}
    S \left( \vec{x}(t_1), \vec{x}(t_2) \right)
    =
    \frac{\vec{x}(t_1) \cdot \vec{x}(t_2)}{\left|\vec{x}(t_1) \right| \left|\vec{x}(t_2) \right|}.
\end{equation}
The temporal decay function to time lag $t$ as shown in Fig.~\ref{fig_RSAandSimilarity} is calculated as 
\begin{equation}
    \phi(t)
    =
    \big\langle S\left( \vec{x}(t_0), \vec{x}(t_0 + t) \right) \big\rangle_{t_0}.
\end{equation}
Since the populations take only positive values, it can relaxe to a finite positive value. Therefore we try fitting $\phi(t)$ by an exponential relaxation to a finite value: $\phi(t) \sim A {\rm e}^{-B t} + (1-A)$ and by a simple power law: $\phi(t) \sim at^{-\lambda t}$. Nevertheless, the simple power law gives better fit to the decay in the high dimensional chaos case. 
Note that, since the populations in each community distribute unevenly as is seen in the instantaneous RSA, it is natural that the decay function in the high-dimensional chaos case relaxes to a value smaller than the expectation value among the population vectors equally distributed on the first orthant of d-dimensional spherical surface, $\left\langle \cos(\theta) \right\rangle_{S^d_+} = d \left[ \frac{2 \Gamma(\frac{d}{2})}{\sqrt{\pi}\Gamma(\frac{d-1}{2})} \right]^2 \xrightarrow[d \to \infty]{} \frac{2}{\pi} \approx 0.6366.$

\section{Supporting Information}

\subsection{Multi attractors}
\begin{figure}[bthp]
\includegraphics[width=1.0\linewidth]{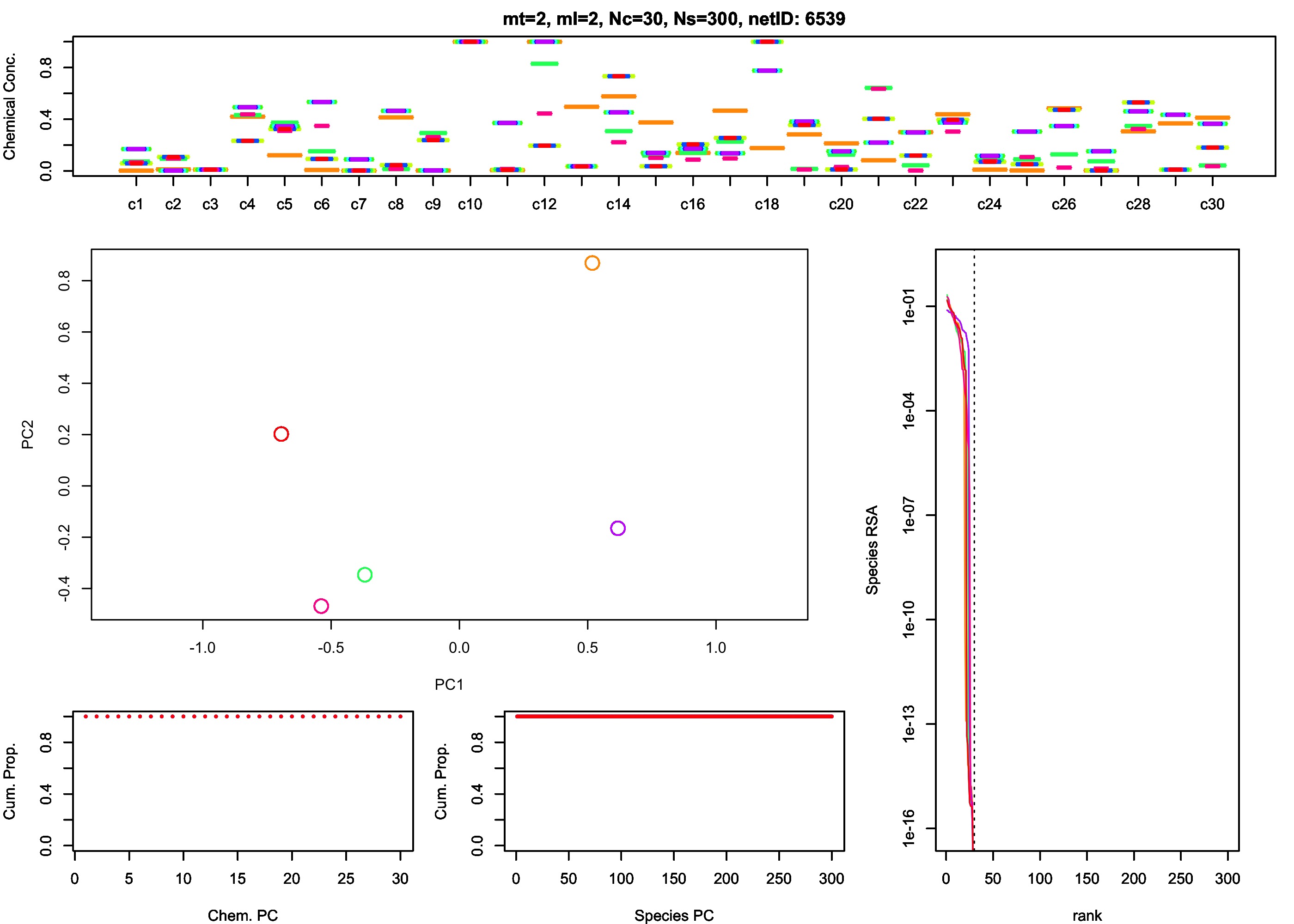}
\caption{An example of multi-fixed points obtained from the same network instance. (Top) The box plot of chemical concentrations $c_i(t)$. (Middle left) Fixed points in the plane of the first and second principal components obtained from of the PCA analysis of sample aggregated chemical dynamics. Note that point size is enlarged for visibility. (Bottom left and center) Cumulative explained-variance ratio plots of the chemical dynamics and population dynamics of each trajectory, respectively. (Bottom right) Relative species abundance plot of cumulative populations of each trajectory.}
\label{SI_fig_multiFP}
\end{figure}

\begin{figure}
\includegraphics[width=1.0\linewidth]{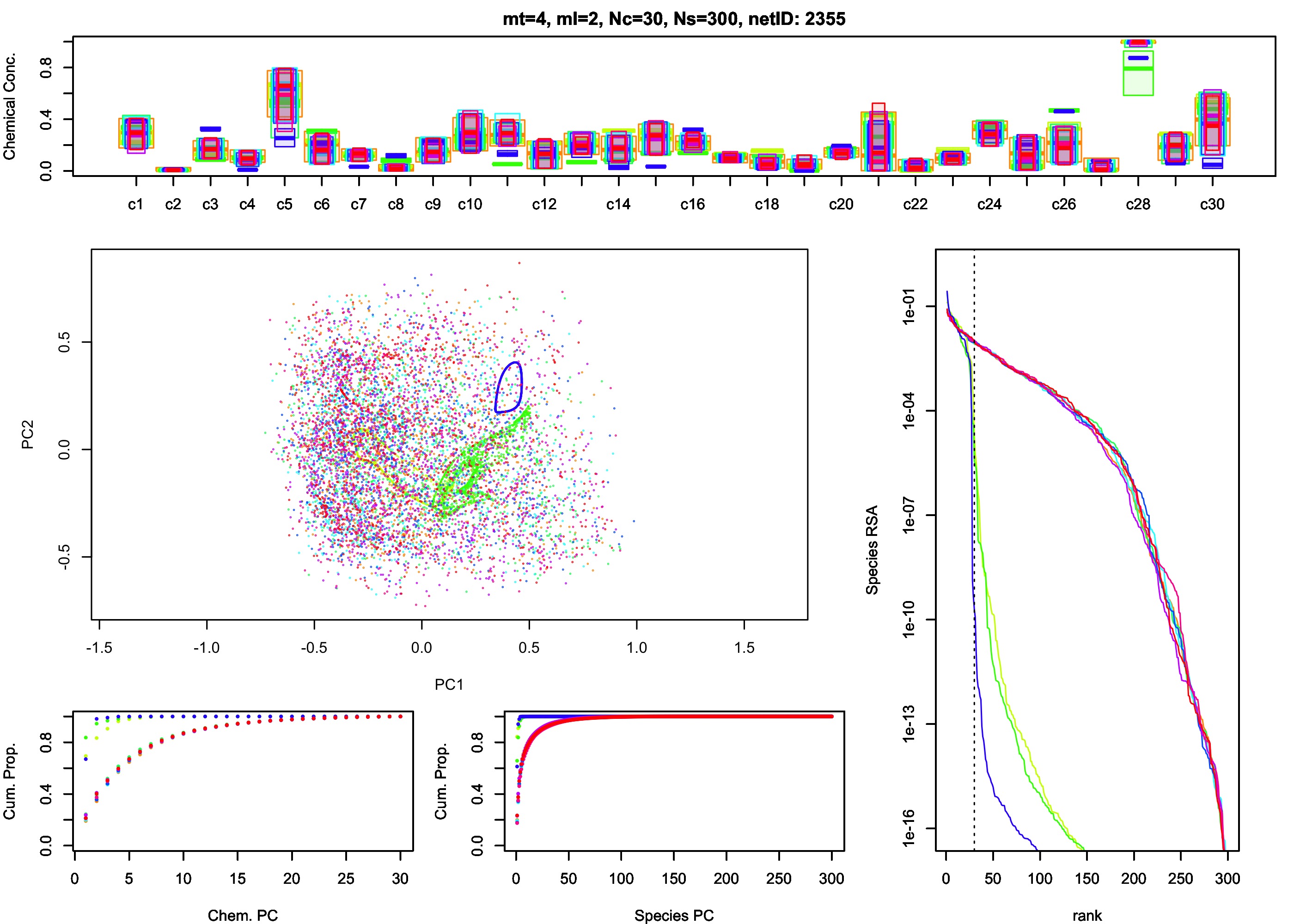}
\caption{An example of multi-attractors obtained from a same network instance: limit cycle, low-dimensional chaos, and high-dimensional chaos. The information shown in each panel is the same as FIG.~\ref{SI_fig_multiFP}.
}
\label{SI_fig_multiAttractor}
\end{figure}

As described in the main text, each network may have multi attractors. FIG.~\ref{SI_fig_multiFP} shows an example of a network which relaxes into different fixed points depending on the initial conditions, resulting the classification of this network into (i). In the case that attractors include different dynamical classes the most complex trajectory is used for characterizing the network. Therefore, the network shown in FIG.~\ref{SI_fig_multiAttractor} is classified as class (iv).

%%%%%%%%%%%%%%%%%%%%%%%%%%%%%%%%%%%%%%%%
\newpage
\subsection{Trajectories of class (ii)}
\begin{figure}[htbp]
\centering
\includegraphics[width=1.0\linewidth]{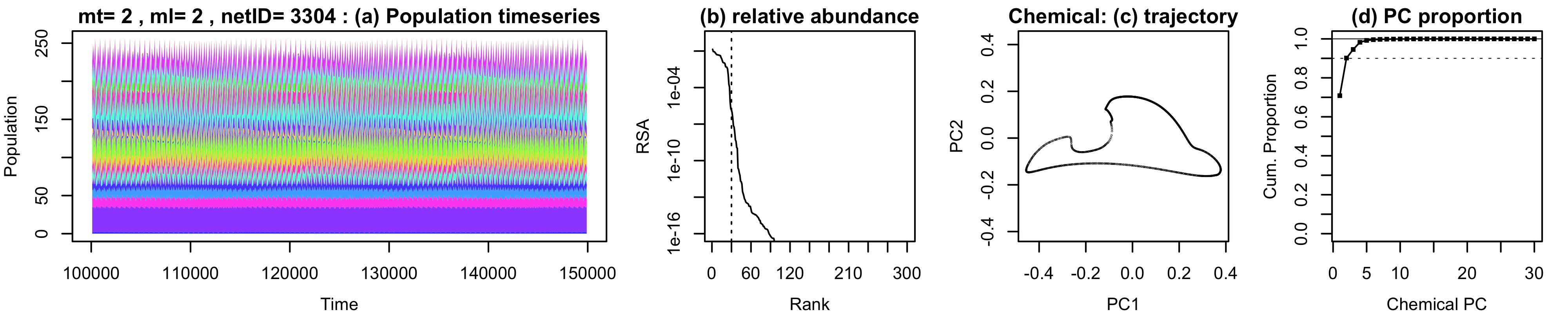}
\includegraphics[width=1.0\linewidth]{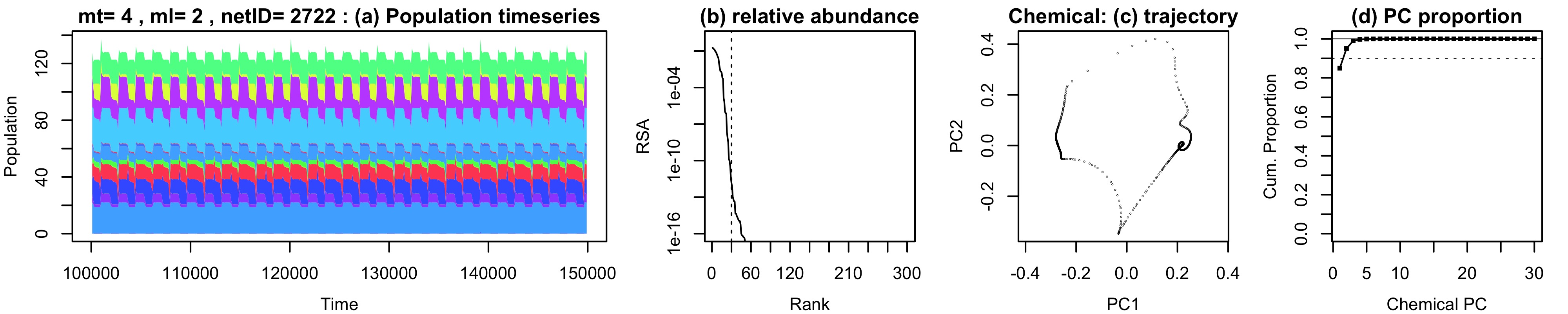}
\includegraphics[width=1.0\linewidth]{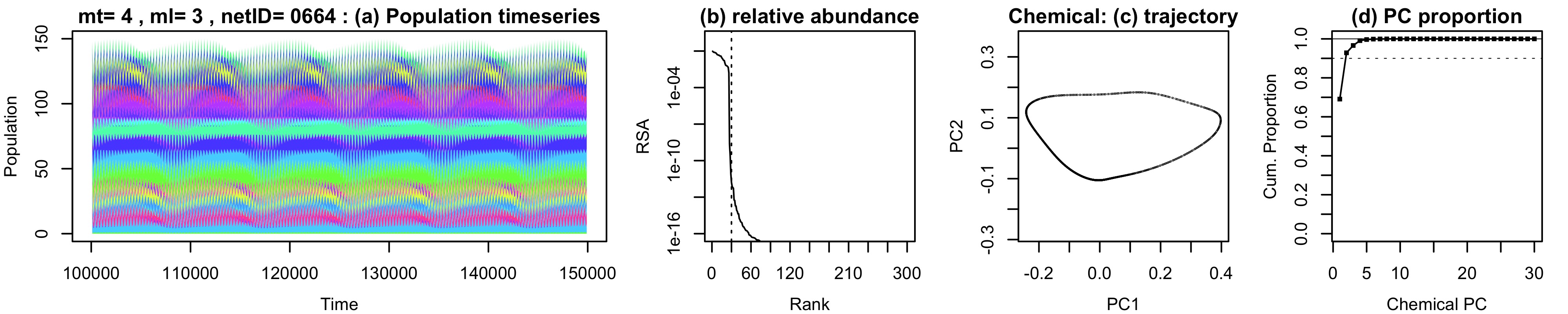}
\includegraphics[width=1.0\linewidth]{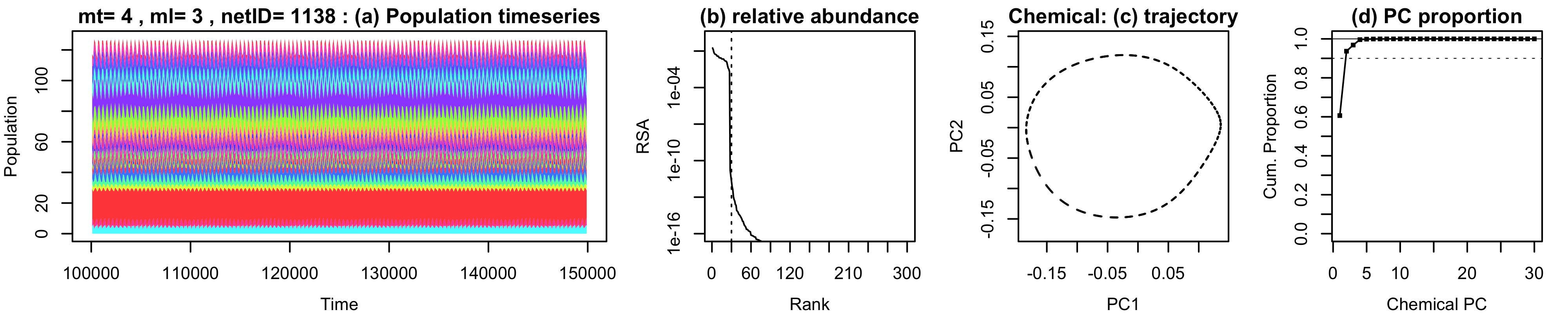}
\includegraphics[width=1.0\linewidth]{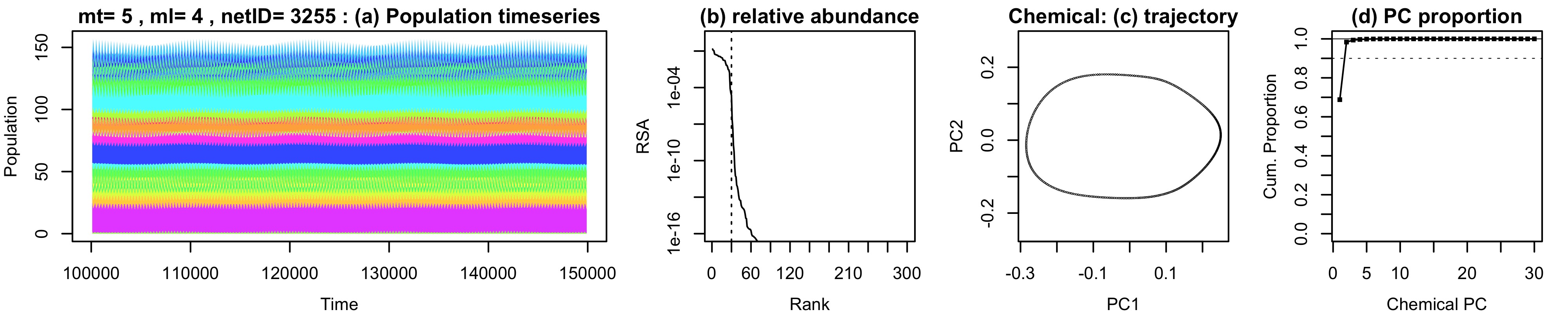}
\caption{Examples of trajectories from the class (ii).}
\label{SI_fig_class2}
\end{figure}

Trajectories classified as class (ii) by its chemical dimension ($1 \ge d_c \ge 2$) indeed corresponds to limit cycles, with diverse amplitude and period, as shown in FIG.~\ref{SI_fig_class2}.

\subsection{Trajectories of class (iii)}
\begin{figure}[htbp]
\centering
\includegraphics[width=1.0\linewidth]{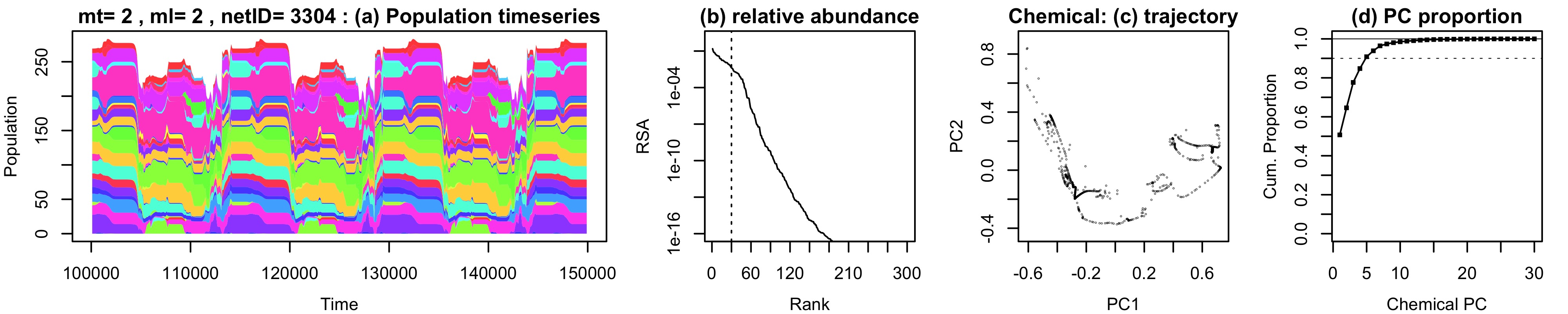}
\includegraphics[width=1.0\linewidth]{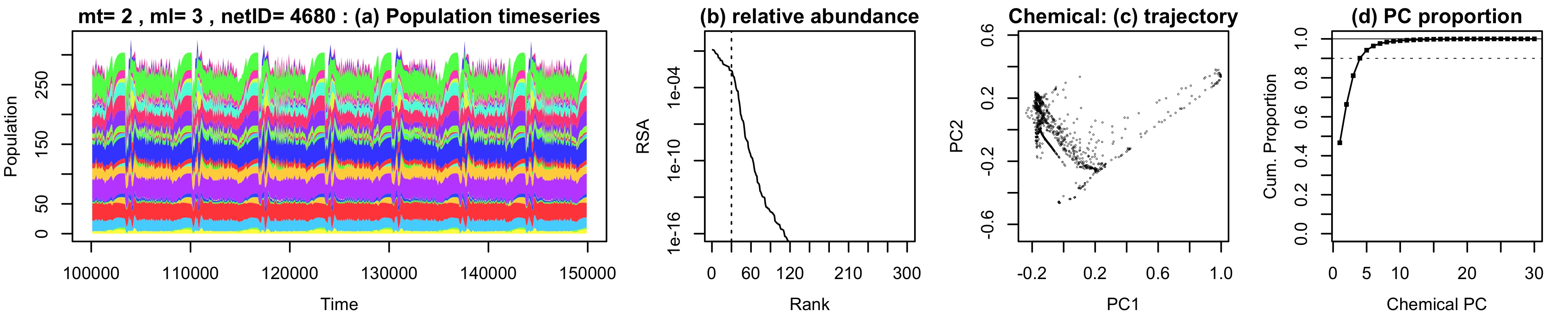}
\includegraphics[width=1.0\linewidth]{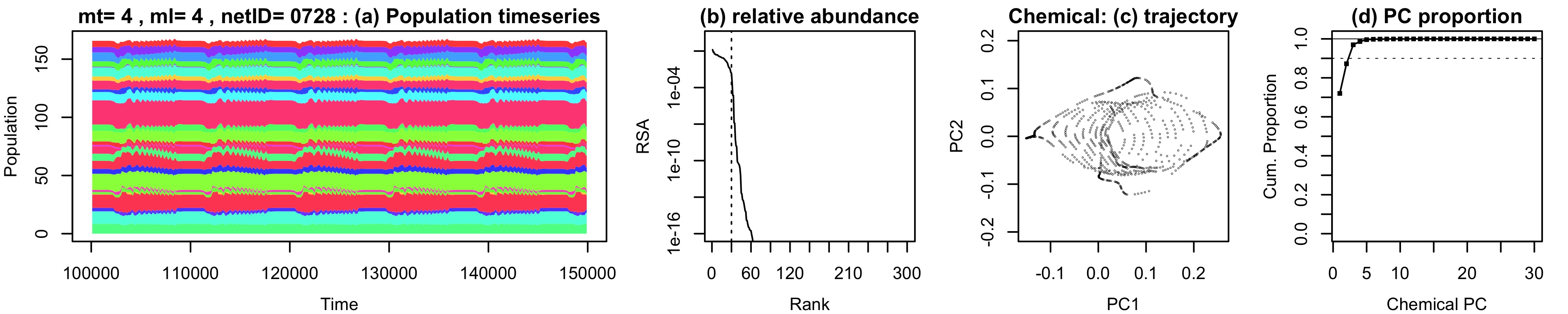}
\includegraphics[width=1.0\linewidth]{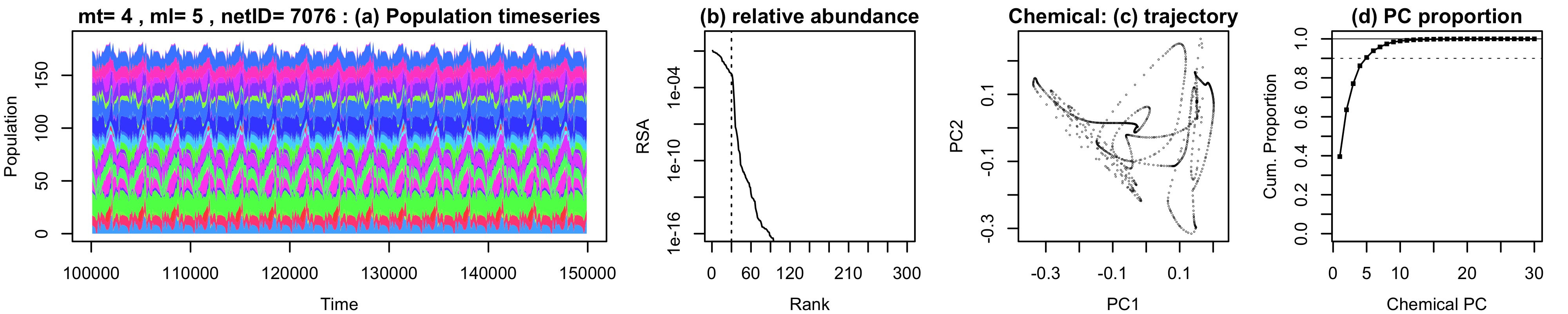}
\includegraphics[width=1.0\linewidth]{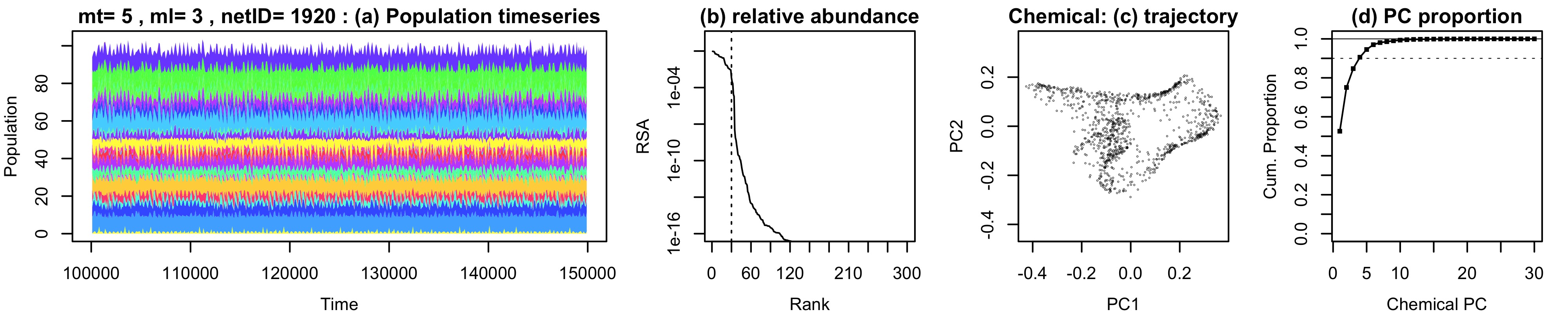}
\caption{Examples of trajectories from the class (iii).}
\label{SI_fig_class3}
\end{figure}

Trajectories classified as class (iii) by its chemical dimension ($3 \ge d_c \ge 7$) include diverse and more complex dynamics including quasi-periodic orbits and low-dimensional chaos, as shown in FIG.~\ref{SI_fig_class3}.

\subsection{Trajectories of class (iv)}
\begin{figure}[htbp]
\centering
\includegraphics[width=1.0\linewidth]{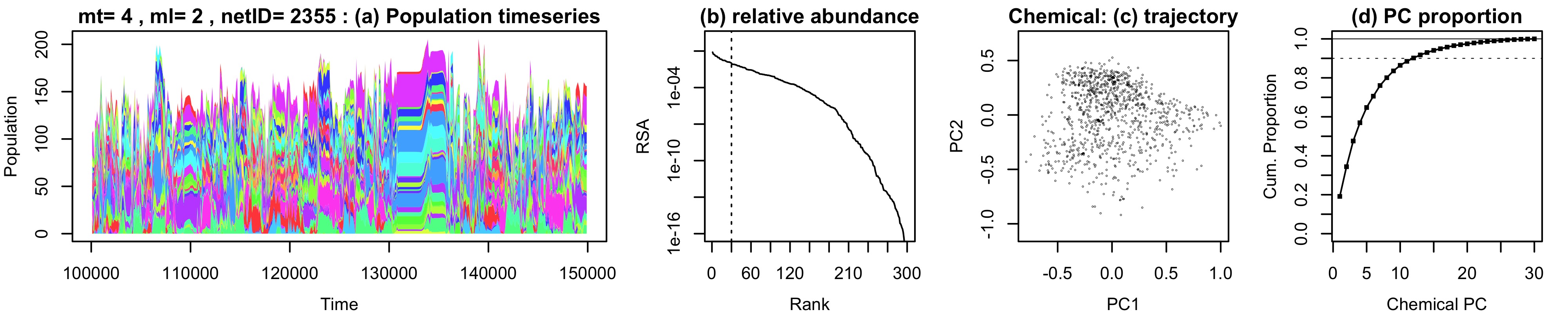}
\includegraphics[width=1.0\linewidth]{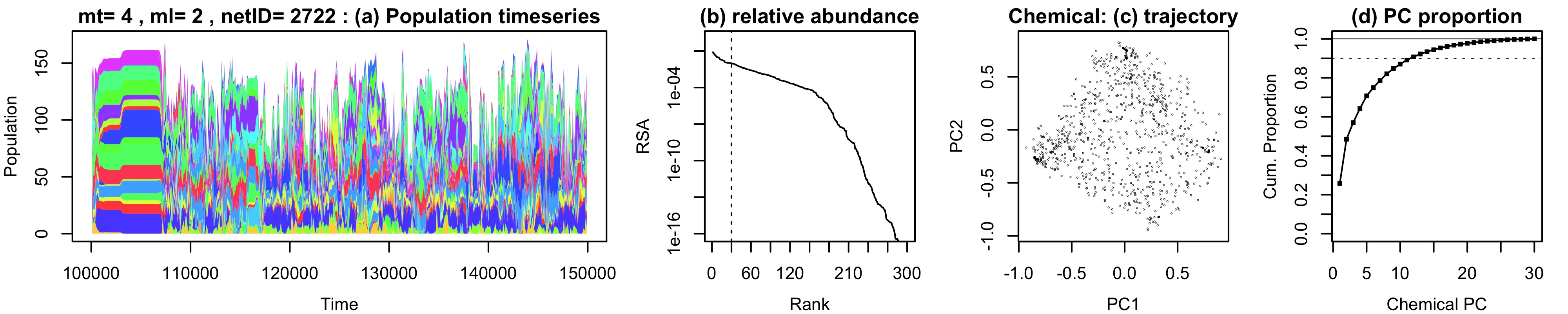}
\includegraphics[width=1.0\linewidth]{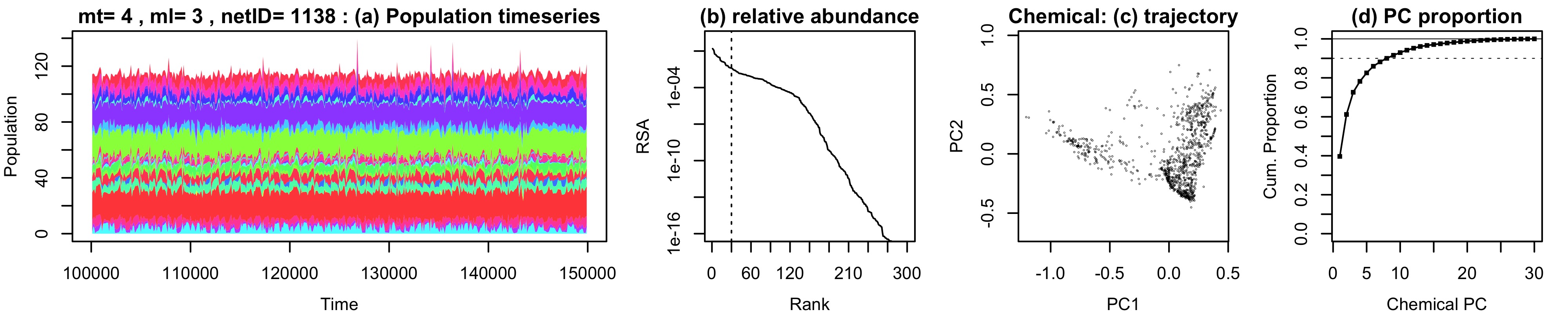}
\includegraphics[width=1.0\linewidth]{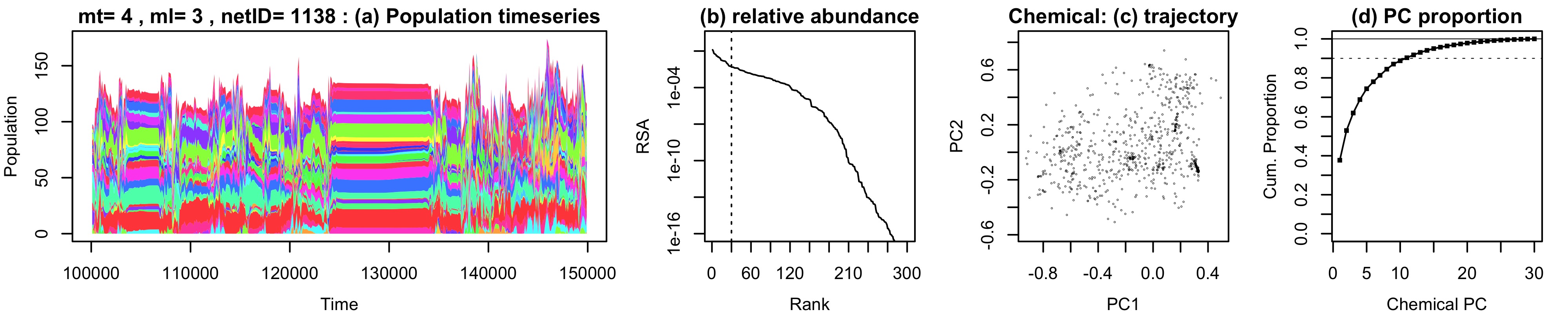}
\includegraphics[width=1.0\linewidth]{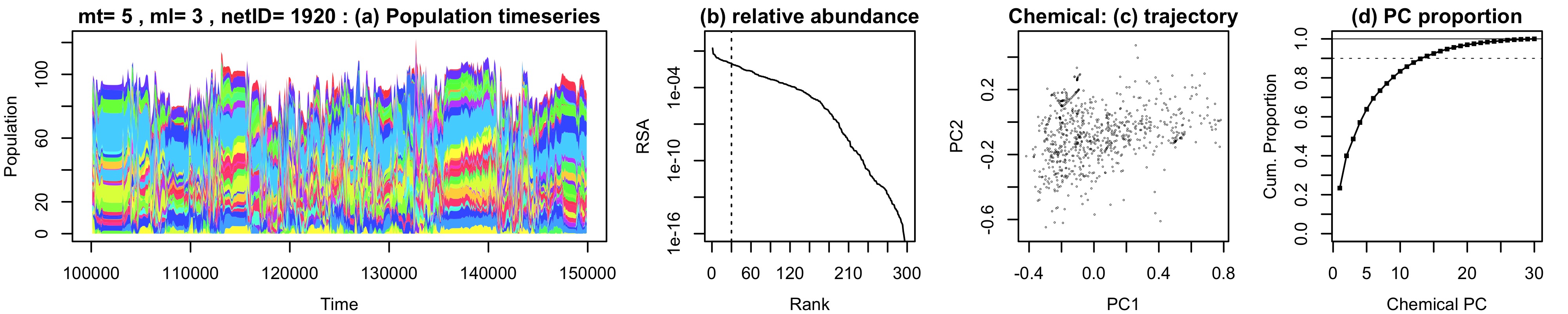}
\caption{Examples of trajectories from the class (iv).}
\label{SI_fig_class4}
\end{figure}

Trajectories classified as class (iv) by its chemical dimension ($8 \ge d_c$) corresponds to high-dimensional chaos. As shown in FIG.~\ref{SI_fig_class4}, the community dynamics shows diverse type of complex behavior including intermittent transitions from an equilibrium to another.

%%%%%%%%%%%%%%%%%%%%%%%%%%%%%%%%%%%%%%%
\subsection{Dependence of the number of coexisting species on the number of initially prepared species}
\begin{figure}[htbp]
\centering
\includegraphics[width=0.8\linewidth]{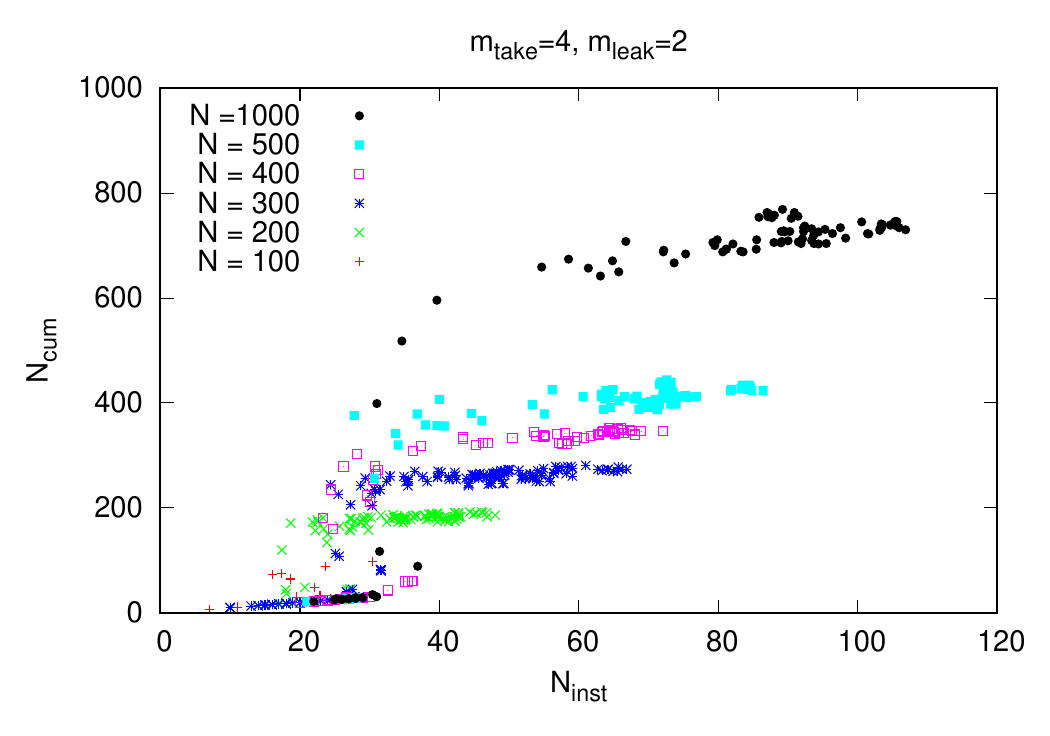}
\caption{$N_{\rm inst}$ and $N_{\rm cum}$ for different $N$, with the same network parameters: $m_{\rm take} = 4, m_{\rm leak} = 2$.}
\label{SI_fig_NdepNcum}
\end{figure}

As shown in FIG.~\ref{SI_fig_NdepNcum}, the ratio of maximum $N_{\rm cum}$ to $N$ for the same network parameters is roughly kept constant. This means that the high-dimensional chaos (iv) covers most dimensions of the entire phase space.

%%%%%%%%%%%%%%%%%%%%%%%%%%%%%%%%%%%%%%%%
\subsection{Dependence of the phase diagram on the number of initially prepared species}
\begin{figure}[htbp]
\centering
\includegraphics[width=0.44\linewidth]{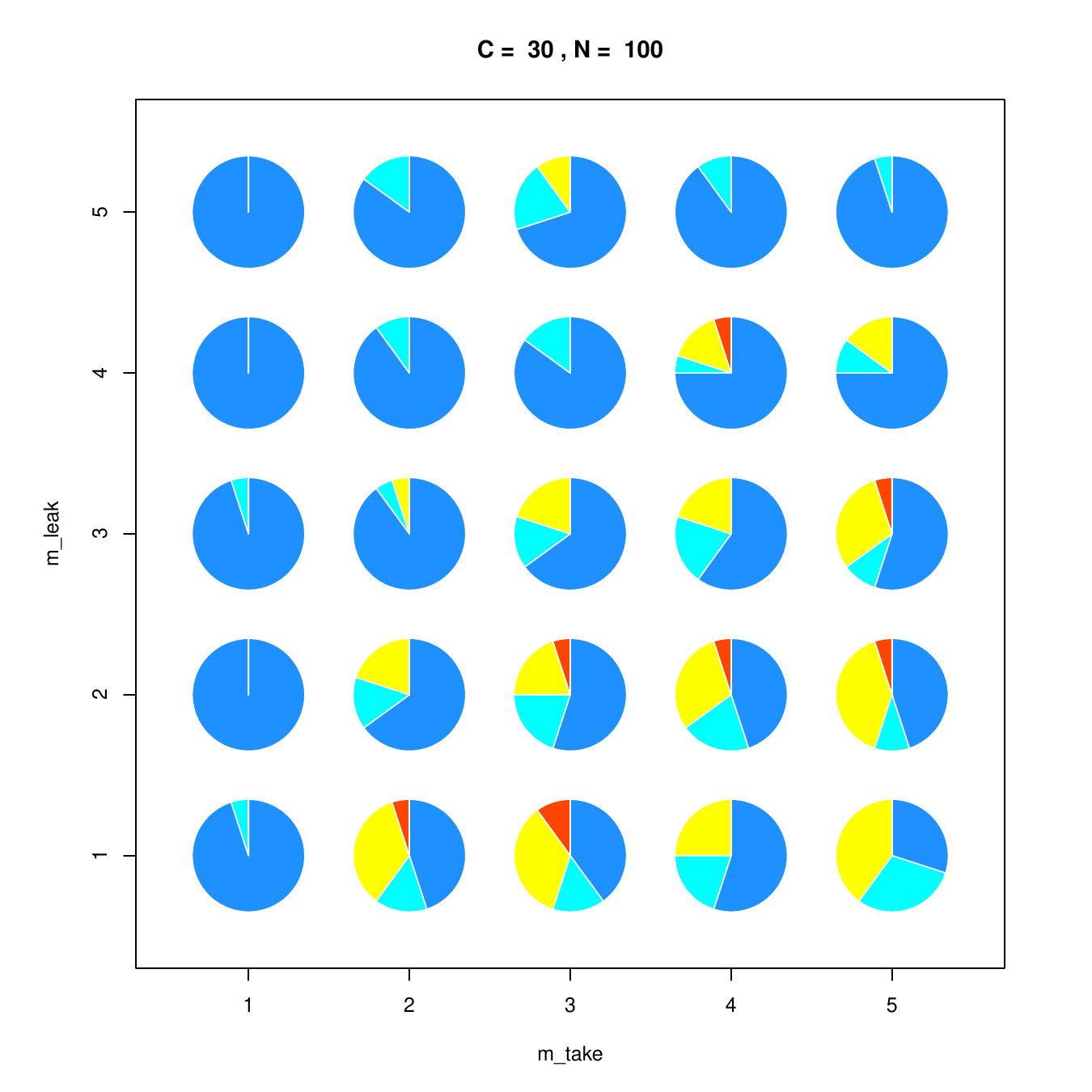}
\includegraphics[width=0.54\linewidth]{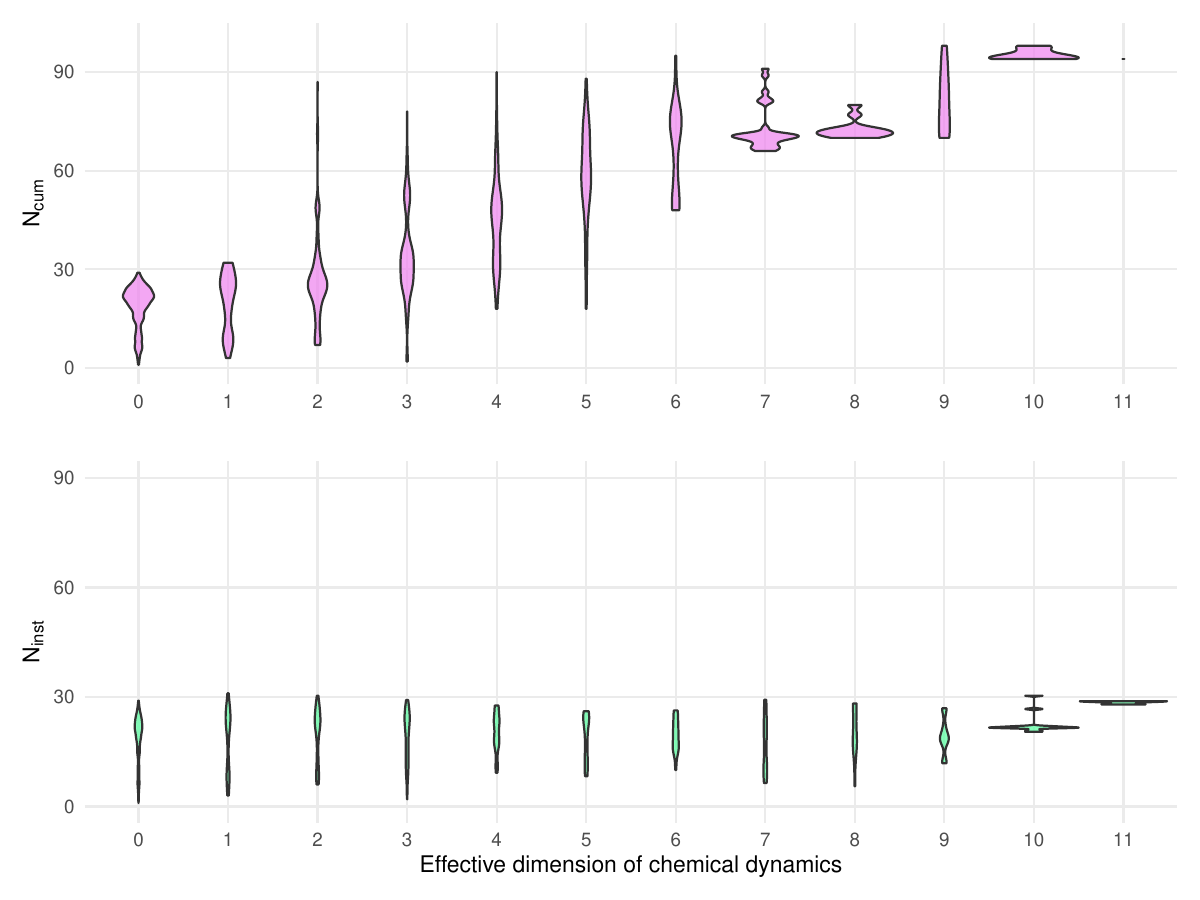}
\includegraphics[width=0.44\linewidth]{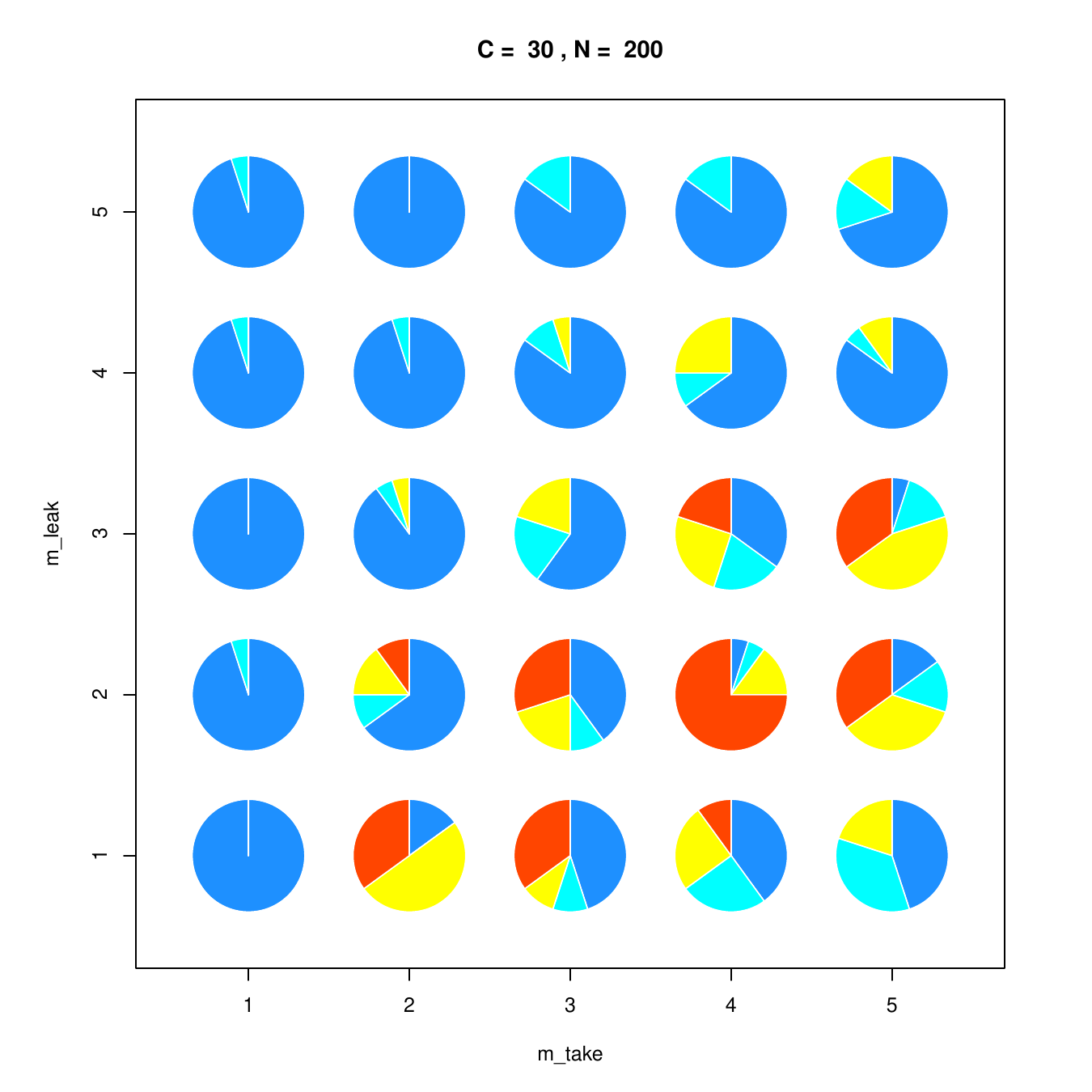}
\includegraphics[width=0.54\linewidth]{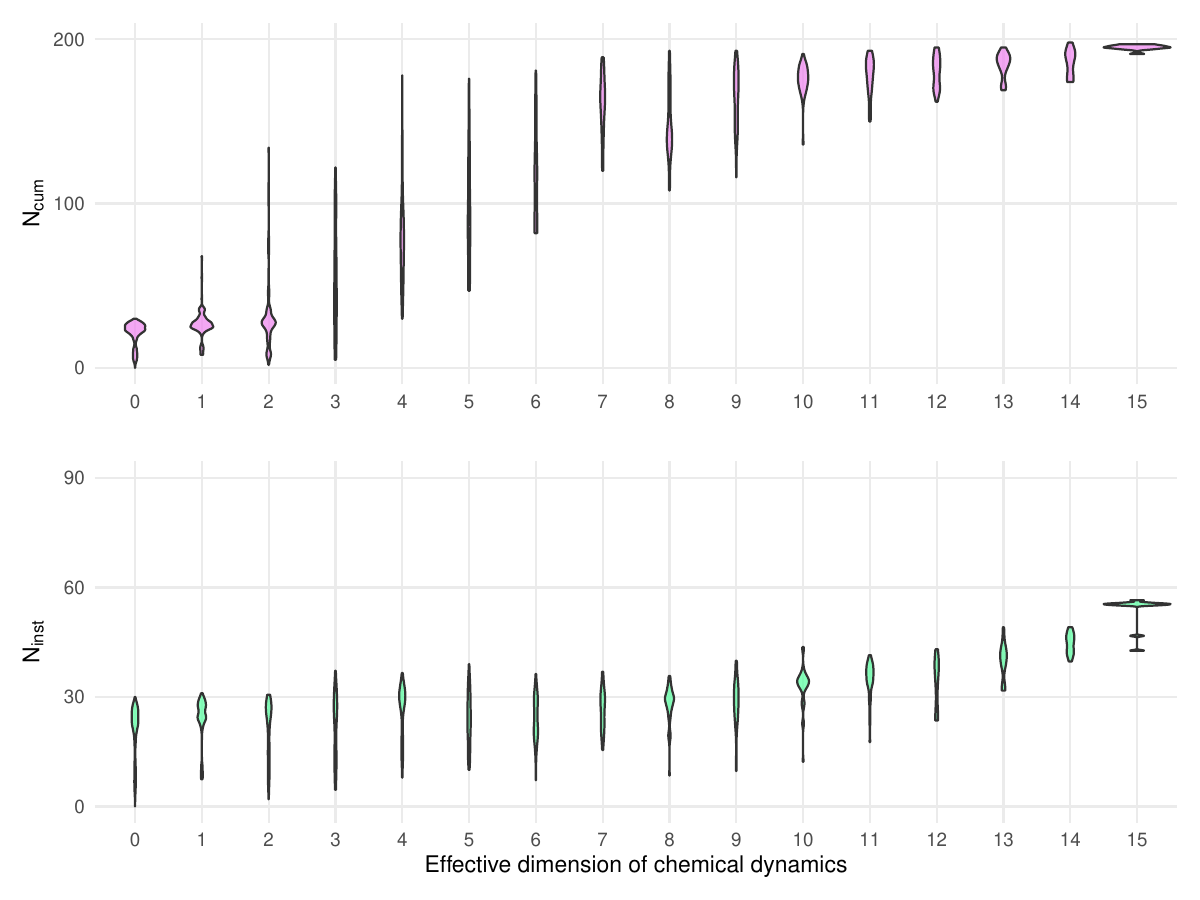}
\includegraphics[width=0.44\linewidth]{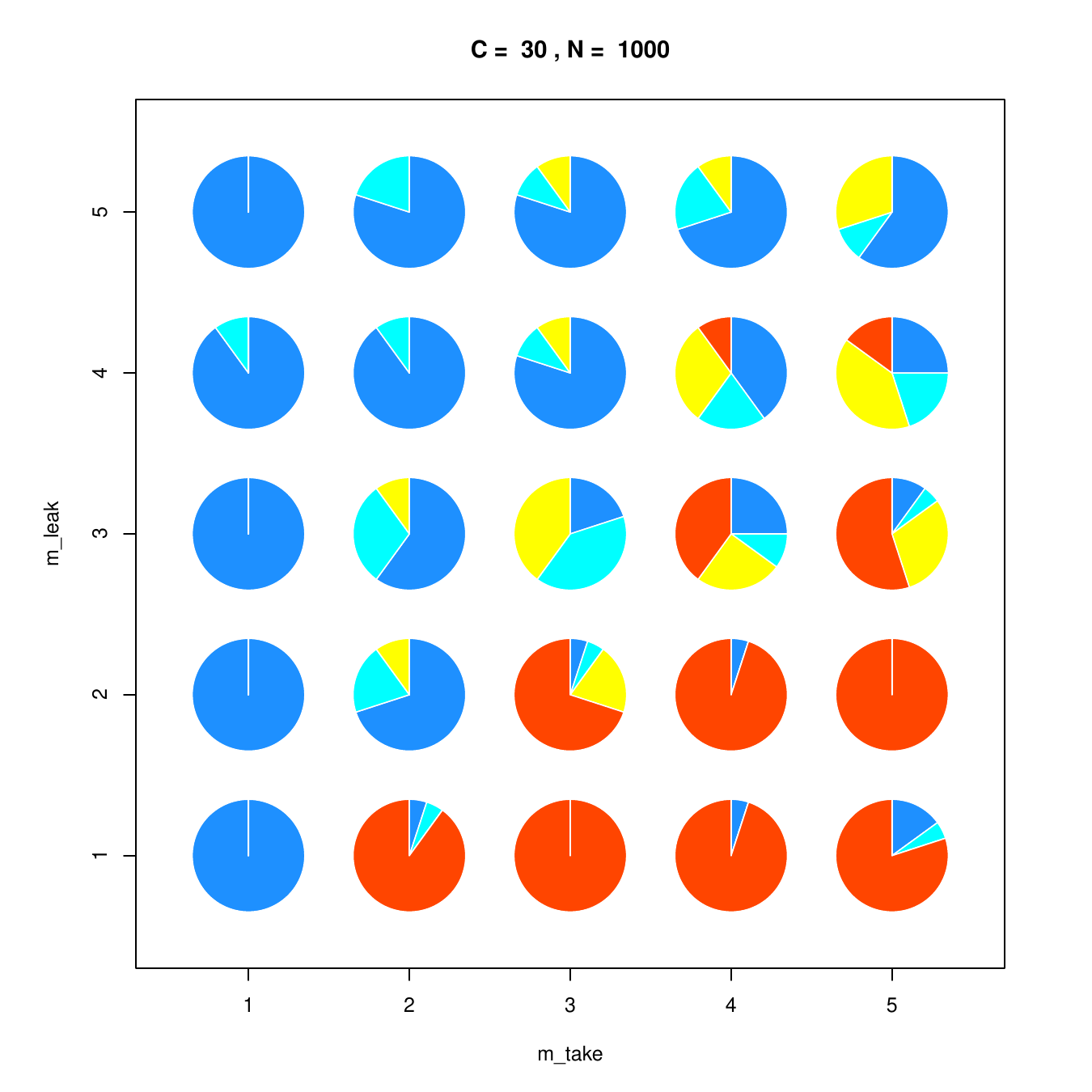}
\includegraphics[width=0.54\linewidth]{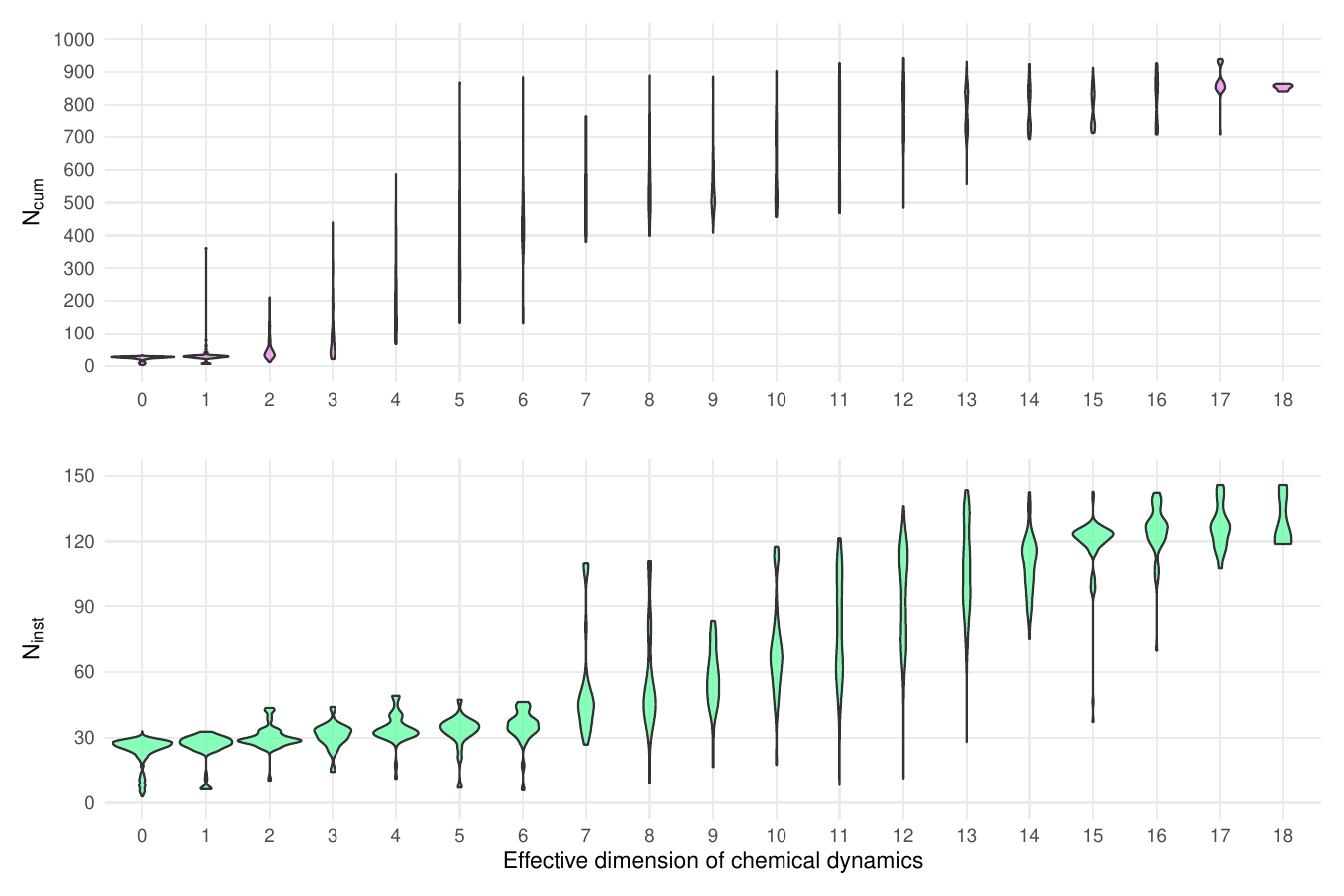}
\caption{The phase diagram of dynamical class and the relation between the dynamical dimension and the coexisting species, in different system sizes, $N=100$ (top), $N=200$ (middle), and $N=1000$ (bottom). The proportion of dynamical classes at each parameter set is calculated from at least $10$ network samples. The result in systems with $N=200$ and $N=1000$ seem to be consistent with the one with $N=300$, in terms of the maxima of $N_{\rm cum} \sim N$ and $D_c \sim 17$. $N=100$ seems to be too small to share this feature.}
\label{SI_fig_NdepPD}
\end{figure}

The number of prepared species $N$ also limit the ``search range'' of the network configuration and hence is relevant to the resulting dynamical classes. As shown in FIG.~\ref{SI_fig_NdepPD}, the probability to have complex dynamics is smaller when $N$ is small. The probability for finding complex dynamics increases with $N$ in the asymmetric regime $m_{take} > m_{leak}$ while it gives almost no change in the other regime, implying that systems in the former regime potentially have configurations to yield complex dynamics.

\subsection{Observation with longer relaxation time}
\begin{figure}[htbp]
\includegraphics[width=0.44\linewidth]{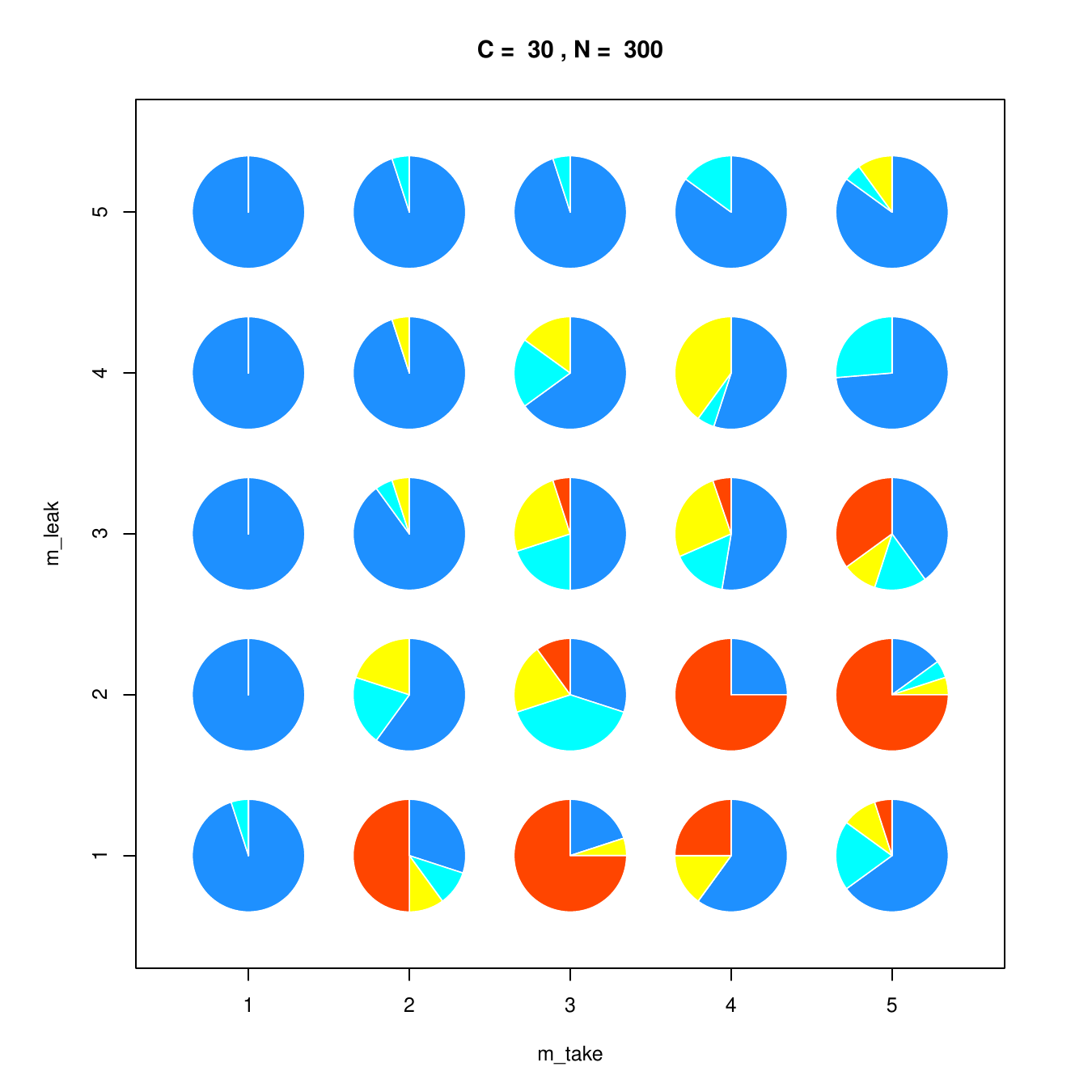}
\includegraphics[width=0.54\linewidth]{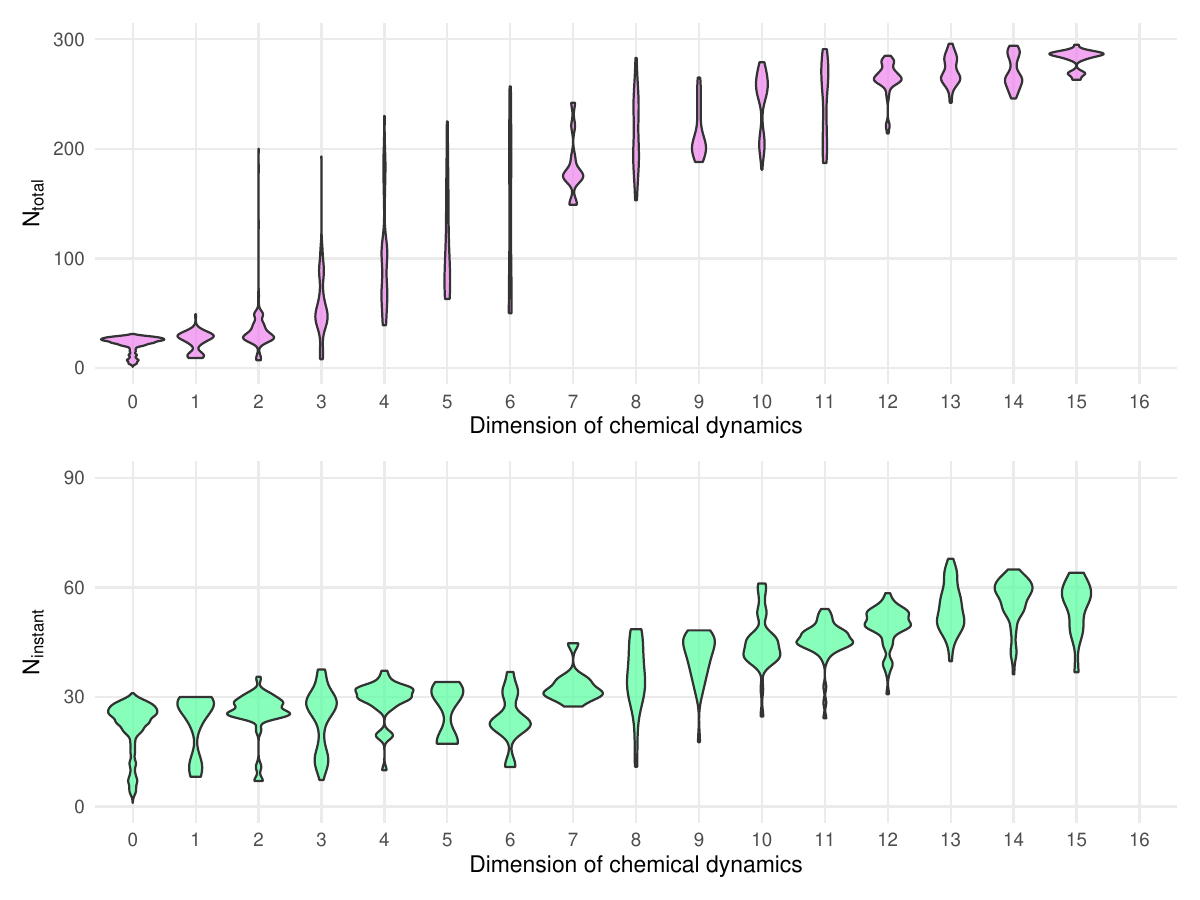}
\caption{Phase diagram obtained from the longer initial relaxation time $T_{\rm ini.} = 4 \times10^5$, instead of $10^5$. System parameters are the same with the main figure: $N = 300, C= 30, b_0 = 100$. This shows that, the low (iii) and high (iv) dimensional chaos argued in the main text include some transient chaos but those are minor or, at least, the transient time is typically very long.}
\label{SI_fig_lrx}
\end{figure}

The dynamics described above may include transient behaviors. Although it is very hard to figure out whether each dynamical behavior is transient or not, we can confirm that the impact of this problem to our conclusion by taking a longer initial relaxation time. As shown in FIG.~\ref{SI_fig_lrx}, the proportion of class (iv) does not show significant decrease against four times longer initial relaxation time. This means that the phase diagram and our main conclusion is not relying on the transient behavior.

\newpage
%%%%%%%%%%%%%%%%%%%%%%%%%%%%%%%%%%%%%%%%%%%%%%%%%%%%%%%%%%%%%%%%%%%%%%%%%%%%
%%%%%%%%%%%%%%%%%%%%%%%%%%%%%%%%%%%%%%%%%%%%%%%%%%%%%%%%%%%%%%%%%%%%%%%%%%%%
% Bibliography
\bibliography{tskk}

%apsrev4-2.bst 2019-01-14 (MD) hand-edited version of apsrev4-1.bst
%Control: key (0)
%Control: author (8) initials jnrlst
%Control: editor formatted (1) identically to author
%Control: production of article title (0) allowed
%Control: page (0) single
%Control: year (1) truncated
%Control: production of eprint (0) enabled
\providecommand{\noopsort}[1]{}\providecommand{\singleletter}[1]{#1}%
\begin{thebibliography}{38}%
\makeatletter
\providecommand \@ifxundefined [1]{%
 \@ifx{#1\undefined}
}%
\providecommand \@ifnum [1]{%
 \ifnum #1\expandafter \@firstoftwo
 \else \expandafter \@secondoftwo
 \fi
}%
\providecommand \@ifx [1]{%
 \ifx #1\expandafter \@firstoftwo
 \else \expandafter \@secondoftwo
 \fi
}%
\providecommand \natexlab [1]{#1}%
\providecommand \enquote  [1]{``#1''}%
\providecommand \bibnamefont  [1]{#1}%
\providecommand \bibfnamefont [1]{#1}%
\providecommand \citenamefont [1]{#1}%
\providecommand \href@noop [0]{\@secondoftwo}%
\providecommand \href [0]{\begingroup \@sanitize@url \@href}%
\providecommand \@href[1]{\@@startlink{#1}\@@href}%
\providecommand \@@href[1]{\endgroup#1\@@endlink}%
\providecommand \@sanitize@url [0]{\catcode `\\12\catcode `\$12\catcode
  `\&12\catcode `\#12\catcode `\^12\catcode `\_12\catcode `\%12\relax}%
\providecommand \@@startlink[1]{}%
\providecommand \@@endlink[0]{}%
\providecommand \url  [0]{\begingroup\@sanitize@url \@url }%
\providecommand \@url [1]{\endgroup\@href {#1}{\urlprefix }}%
\providecommand \urlprefix  [0]{URL }%
\providecommand \Eprint [0]{\href }%
\providecommand \doibase [0]{https://doi.org/}%
\providecommand \selectlanguage [0]{\@gobble}%
\providecommand \bibinfo  [0]{\@secondoftwo}%
\providecommand \bibfield  [0]{\@secondoftwo}%
\providecommand \translation [1]{[#1]}%
\providecommand \BibitemOpen [0]{}%
\providecommand \bibitemStop [0]{}%
\providecommand \bibitemNoStop [0]{.\EOS\space}%
\providecommand \EOS [0]{\spacefactor3000\relax}%
\providecommand \BibitemShut  [1]{\csname bibitem#1\endcsname}%
\let\auto@bib@innerbib\@empty
%</preamble>
\bibitem [{\citenamefont {Macarthur}\ and\ \citenamefont
  {Levins}(1967)}]{MacArthur_AmNat1967}%
  \BibitemOpen
  \bibfield  {author} {\bibinfo {author} {\bibfnamefont {R.}~\bibnamefont
  {Macarthur}}\ and\ \bibinfo {author} {\bibfnamefont {R.}~\bibnamefont
  {Levins}},\ }\bibfield  {title} {\bibinfo {title} {The limiting similarity,
  convergence, and divergence of coexisting species},\ }\href
  {http://www.jstor.org/stable/2459090} {\bibfield  {journal} {\bibinfo
  {journal} {The American Naturalist}\ }\textbf {\bibinfo {volume} {101}},\
  \bibinfo {pages} {377} (\bibinfo {year} {1967})}\BibitemShut {NoStop}%
\bibitem [{\citenamefont {Cui}\ \emph {et~al.}(2020)\citenamefont {Cui},
  \citenamefont {Marsland},\ and\ \citenamefont
  {Mehta}}]{CuiMarslandMetha_PRL2020}%
  \BibitemOpen
  \bibfield  {author} {\bibinfo {author} {\bibfnamefont {W.}~\bibnamefont
  {Cui}}, \bibinfo {author} {\bibfnamefont {R.}~\bibnamefont {Marsland}},\ and\
  \bibinfo {author} {\bibfnamefont {P.}~\bibnamefont {Mehta}},\ }\bibfield
  {title} {\bibinfo {title} {Effect of resource dynamics on species packing in
  diverse ecosystems},\ }\href {https://doi.org/10.1103/PhysRevLett.125.048101}
  {\bibfield  {journal} {\bibinfo  {journal} {Phys. Rev. Lett.}\ }\textbf
  {\bibinfo {volume} {125}},\ \bibinfo {pages} {048101} (\bibinfo {year}
  {2020})}\BibitemShut {NoStop}%
\bibitem [{\citenamefont {Moran}\ and\ \citenamefont
  {Tikhonov}(2022)}]{MoranTikhonov_PhysRevX2022}%
  \BibitemOpen
  \bibfield  {author} {\bibinfo {author} {\bibfnamefont {J.}~\bibnamefont
  {Moran}}\ and\ \bibinfo {author} {\bibfnamefont {M.}~\bibnamefont
  {Tikhonov}},\ }\bibfield  {title} {\bibinfo {title} {Defining
  coarse-grainability in a model of structured microbial ecosystems},\ }\href
  {https://doi.org/10.1103/PhysRevX.12.021038} {\bibfield  {journal} {\bibinfo
  {journal} {Phys. Rev. X}\ }\textbf {\bibinfo {volume} {12}},\ \bibinfo
  {pages} {021038} (\bibinfo {year} {2022})}\BibitemShut {NoStop}%
\bibitem [{\citenamefont {Blumenthal}\ \emph {et~al.}(2024)\citenamefont
  {Blumenthal}, \citenamefont {Rocks},\ and\ \citenamefont
  {Mehta}}]{BlumenthalMetha_PRL2024}%
  \BibitemOpen
  \bibfield  {author} {\bibinfo {author} {\bibfnamefont {E.}~\bibnamefont
  {Blumenthal}}, \bibinfo {author} {\bibfnamefont {J.~W.}\ \bibnamefont
  {Rocks}},\ and\ \bibinfo {author} {\bibfnamefont {P.}~\bibnamefont {Mehta}},\
  }\bibfield  {title} {\bibinfo {title} {Phase transition to chaos in complex
  ecosystems with nonreciprocal species-resource interactions},\ }\href
  {https://doi.org/10.1103/PhysRevLett.132.127401} {\bibfield  {journal}
  {\bibinfo  {journal} {Phys. Rev. Lett.}\ }\textbf {\bibinfo {volume} {132}},\
  \bibinfo {pages} {127401} (\bibinfo {year} {2024})}\BibitemShut {NoStop}%
\bibitem [{\citenamefont {Gause}(1932)}]{Gause1932}%
  \BibitemOpen
  \bibfield  {author} {\bibinfo {author} {\bibfnamefont {G.~F.}\ \bibnamefont
  {Gause}},\ }\bibfield  {title} {\bibinfo {title} {Experimental studies on the
  struggle for existence: I. mixed population of two species of yeast},\ }\href
  {https://doi.org/10.1242/jeb.9.4.389} {\bibfield  {journal} {\bibinfo
  {journal} {Journal of Experimental Biology}\ }\textbf {\bibinfo {volume}
  {9}},\ \bibinfo {pages} {389} (\bibinfo {year} {1932})}\BibitemShut {NoStop}%
\bibitem [{\citenamefont {Hardin}(1960)}]{Hardin1960}%
  \BibitemOpen
  \bibfield  {author} {\bibinfo {author} {\bibfnamefont {G.}~\bibnamefont
  {Hardin}},\ }\bibfield  {title} {\bibinfo {title} {The competitive exclusion
  principle.},\ }\href {https://doi.org/10.1126/science.131.3409.1292}
  {\bibfield  {journal} {\bibinfo  {journal} {Science}\ }\textbf {\bibinfo
  {volume} {131}},\ \bibinfo {pages} {1292} (\bibinfo {year}
  {1960})}\BibitemShut {NoStop}%
\bibitem [{\citenamefont {Hutchinson}(1961)}]{Hutchinson_AmNat1961}%
  \BibitemOpen
  \bibfield  {author} {\bibinfo {author} {\bibfnamefont {G.~E.}\ \bibnamefont
  {Hutchinson}},\ }\bibfield  {title} {\bibinfo {title} {The paradox of the
  plankton},\ }\href {https://doi.org/10.1086/282171} {\bibfield  {journal}
  {\bibinfo  {journal} {The American Naturalist}\ }\textbf {\bibinfo {volume}
  {95}},\ \bibinfo {pages} {137} (\bibinfo {year} {1961})},\ \Eprint
  {https://arxiv.org/abs/https://doi.org/10.1086/282171}
  {https://doi.org/10.1086/282171} \BibitemShut {NoStop}%
\bibitem [{\citenamefont {Datta}\ \emph {et~al.}(2016)\citenamefont {Datta},
  \citenamefont {Sliwerska}, \citenamefont {Gore}, \citenamefont {Polz},\ and\
  \citenamefont {Cordero}}]{DattaCordero_NatComm2016}%
  \BibitemOpen
  \bibfield  {author} {\bibinfo {author} {\bibfnamefont {M.~S.}\ \bibnamefont
  {Datta}}, \bibinfo {author} {\bibfnamefont {E.}~\bibnamefont {Sliwerska}},
  \bibinfo {author} {\bibfnamefont {J.}~\bibnamefont {Gore}}, \bibinfo {author}
  {\bibfnamefont {M.~F.}\ \bibnamefont {Polz}},\ and\ \bibinfo {author}
  {\bibfnamefont {O.~X.}\ \bibnamefont {Cordero}},\ }\bibfield  {title}
  {\bibinfo {title} {Microbial interactions lead to rapid micro-scale
  successions on model marine particles},\ }\href
  {https://doi.org/10.1038/ncomms11965} {\bibfield  {journal} {\bibinfo
  {journal} {Nature Communications}\ }\textbf {\bibinfo {volume} {7}},\
  \bibinfo {pages} {11965} (\bibinfo {year} {2016})}\BibitemShut {NoStop}%
\bibitem [{\citenamefont {Zelezniak}\ \emph {et~al.}(2015)\citenamefont
  {Zelezniak}, \citenamefont {Andrejev}, \citenamefont {Ponomarova},
  \citenamefont {Mende}, \citenamefont {Bork},\ and\ \citenamefont
  {Patil}}]{ZelezniakPatil_PNAS2015}%
  \BibitemOpen
  \bibfield  {author} {\bibinfo {author} {\bibfnamefont {A.}~\bibnamefont
  {Zelezniak}}, \bibinfo {author} {\bibfnamefont {S.}~\bibnamefont {Andrejev}},
  \bibinfo {author} {\bibfnamefont {O.}~\bibnamefont {Ponomarova}}, \bibinfo
  {author} {\bibfnamefont {D.~R.}\ \bibnamefont {Mende}}, \bibinfo {author}
  {\bibfnamefont {P.}~\bibnamefont {Bork}},\ and\ \bibinfo {author}
  {\bibfnamefont {K.~R.}\ \bibnamefont {Patil}},\ }\bibfield  {title} {\bibinfo
  {title} {Metabolic dependencies drive species co-occurrence in diverse
  microbial communities},\ }\href {https://doi.org/10.1073/pnas.1421834112}
  {\bibfield  {journal} {\bibinfo  {journal} {Proceedings of the National
  Academy of Sciences}\ }\textbf {\bibinfo {volume} {112}},\ \bibinfo {pages}
  {6449} (\bibinfo {year} {2015})},\ \Eprint
  {https://arxiv.org/abs/https://www.pnas.org/doi/pdf/10.1073/pnas.1421834112}
  {https://www.pnas.org/doi/pdf/10.1073/pnas.1421834112} \BibitemShut {NoStop}%
\bibitem [{\citenamefont {Goyal}\ and\ \citenamefont
  {Maslov}(2018)}]{GoyalMasrov_PRL2018}%
  \BibitemOpen
  \bibfield  {author} {\bibinfo {author} {\bibfnamefont {A.}~\bibnamefont
  {Goyal}}\ and\ \bibinfo {author} {\bibfnamefont {S.}~\bibnamefont {Maslov}},\
  }\bibfield  {title} {\bibinfo {title} {Diversity, stability, and
  reproducibility in stochastically assembled microbial ecosystems},\ }\href
  {https://doi.org/10.1103/PhysRevLett.120.158102} {\bibfield  {journal}
  {\bibinfo  {journal} {Phys. Rev. Lett.}\ }\textbf {\bibinfo {volume} {120}},\
  \bibinfo {pages} {158102} (\bibinfo {year} {2018})}\BibitemShut {NoStop}%
\bibitem [{\citenamefont {Goldford}\ \emph {et~al.}(2018)\citenamefont
  {Goldford}, \citenamefont {Lu}, \citenamefont {Baji\'{c}}, \citenamefont
  {Estrela}, \citenamefont {Tikhonov}, \citenamefont {Sanchez-Gorostiaga},
  \citenamefont {Segr\`{e}}, \citenamefont {Mehta},\ and\ \citenamefont
  {Sanchez}}]{GoldfordTikhonovMehtaSanchez_Science2018}%
  \BibitemOpen
  \bibfield  {author} {\bibinfo {author} {\bibfnamefont {J.~E.}\ \bibnamefont
  {Goldford}}, \bibinfo {author} {\bibfnamefont {N.}~\bibnamefont {Lu}},
  \bibinfo {author} {\bibfnamefont {D.}~\bibnamefont {Baji\'{c}}}, \bibinfo
  {author} {\bibfnamefont {S.}~\bibnamefont {Estrela}}, \bibinfo {author}
  {\bibfnamefont {M.}~\bibnamefont {Tikhonov}}, \bibinfo {author}
  {\bibfnamefont {A.}~\bibnamefont {Sanchez-Gorostiaga}}, \bibinfo {author}
  {\bibfnamefont {D.}~\bibnamefont {Segr\`{e}}}, \bibinfo {author}
  {\bibfnamefont {P.}~\bibnamefont {Mehta}},\ and\ \bibinfo {author}
  {\bibfnamefont {A.}~\bibnamefont {Sanchez}},\ }\bibfield  {title} {\bibinfo
  {title} {Emergent simplicity in microbial community assembly},\ }\href
  {https://doi.org/10.1126/science.aat1168} {\bibfield  {journal} {\bibinfo
  {journal} {Science}\ }\textbf {\bibinfo {volume} {361}},\ \bibinfo {pages}
  {469} (\bibinfo {year} {2018})},\ \Eprint
  {https://arxiv.org/abs/https://www.science.org/doi/pdf/10.1126/science.aat1168}
  {https://www.science.org/doi/pdf/10.1126/science.aat1168} \BibitemShut
  {NoStop}%
\bibitem [{\citenamefont {D{'}Souza}\ \emph {et~al.}(2018)\citenamefont
  {D{'}Souza}, \citenamefont {Shitut}, \citenamefont {Preussger}, \citenamefont
  {Yousif}, \citenamefont {Waschina},\ and\ \citenamefont
  {Kost}}]{DSouzaKost_NatProdRep2018}%
  \BibitemOpen
  \bibfield  {author} {\bibinfo {author} {\bibfnamefont {G.}~\bibnamefont
  {D{'}Souza}}, \bibinfo {author} {\bibfnamefont {S.}~\bibnamefont {Shitut}},
  \bibinfo {author} {\bibfnamefont {D.}~\bibnamefont {Preussger}}, \bibinfo
  {author} {\bibfnamefont {G.}~\bibnamefont {Yousif}}, \bibinfo {author}
  {\bibfnamefont {S.}~\bibnamefont {Waschina}},\ and\ \bibinfo {author}
  {\bibfnamefont {C.}~\bibnamefont {Kost}},\ }\bibfield  {title} {\bibinfo
  {title} {Ecology and evolution of metabolic cross-feeding interactions in
  bacteria},\ }\href {https://doi.org/10.1039/C8NP00009C} {\bibfield  {journal}
  {\bibinfo  {journal} {Nat. Prod. Rep.}\ }\textbf {\bibinfo {volume} {35}},\
  \bibinfo {pages} {455} (\bibinfo {year} {2018})}\BibitemShut {NoStop}%
\bibitem [{\citenamefont {Rosenzweig}\ \emph {et~al.}(1994)\citenamefont
  {Rosenzweig}, \citenamefont {Sharp}, \citenamefont {Treves},\ and\
  \citenamefont {Adams}}]{RosenzweigAdams_Genetics1994}%
  \BibitemOpen
  \bibfield  {author} {\bibinfo {author} {\bibfnamefont {R.~F.}\ \bibnamefont
  {Rosenzweig}}, \bibinfo {author} {\bibfnamefont {R.~R.}\ \bibnamefont
  {Sharp}}, \bibinfo {author} {\bibfnamefont {D.~S.}\ \bibnamefont {Treves}},\
  and\ \bibinfo {author} {\bibfnamefont {J.}~\bibnamefont {Adams}},\ }\bibfield
   {title} {\bibinfo {title} {Microbial evolution in a simple unstructured
  environment: genetic differentiation in escherichia coli.},\ }\href
  {https://doi.org/10.1093/genetics/137.4.903} {\bibfield  {journal} {\bibinfo
  {journal} {Genetics}\ }\textbf {\bibinfo {volume} {137}},\ \bibinfo {pages}
  {903} (\bibinfo {year} {1994})},\ \Eprint
  {https://arxiv.org/abs/https://academic.oup.com/genetics/article-pdf/137/4/903/37005764/genetics0903.pdf}
  {https://academic.oup.com/genetics/article-pdf/137/4/903/37005764/genetics0903.pdf}
  \BibitemShut {NoStop}%
\bibitem [{\citenamefont {Ponomarova}\ \emph {et~al.}(2017)\citenamefont
  {Ponomarova}, \citenamefont {Gabrielli}, \citenamefont {S\'{e}vin},
  \citenamefont {M\"{u}lleder}, \citenamefont {Zirngibl}, \citenamefont
  {Bulyha}, \citenamefont {Andrejev}, \citenamefont {Kafkia}, \citenamefont
  {Typas}, \citenamefont {Sauer}, \citenamefont {Ralser},\ and\ \citenamefont
  {Patil}}]{PonomarovaPatil_CellSystems2017}%
  \BibitemOpen
  \bibfield  {author} {\bibinfo {author} {\bibfnamefont {O.}~\bibnamefont
  {Ponomarova}}, \bibinfo {author} {\bibfnamefont {N.}~\bibnamefont
  {Gabrielli}}, \bibinfo {author} {\bibfnamefont {D.~C.}\ \bibnamefont
  {S\'{e}vin}}, \bibinfo {author} {\bibfnamefont {M.}~\bibnamefont
  {M\"{u}lleder}}, \bibinfo {author} {\bibfnamefont {K.}~\bibnamefont
  {Zirngibl}}, \bibinfo {author} {\bibfnamefont {K.}~\bibnamefont {Bulyha}},
  \bibinfo {author} {\bibfnamefont {S.}~\bibnamefont {Andrejev}}, \bibinfo
  {author} {\bibfnamefont {E.}~\bibnamefont {Kafkia}}, \bibinfo {author}
  {\bibfnamefont {A.}~\bibnamefont {Typas}}, \bibinfo {author} {\bibfnamefont
  {U.}~\bibnamefont {Sauer}}, \bibinfo {author} {\bibfnamefont
  {M.}~\bibnamefont {Ralser}},\ and\ \bibinfo {author} {\bibfnamefont {K.~R.}\
  \bibnamefont {Patil}},\ }\bibfield  {title} {\bibinfo {title} {Yeast creates
  a niche for symbiotic lactic acid bacteria through nitrogen overflow},\
  }\href {https://doi.org/https://doi.org/10.1016/j.cels.2017.09.002}
  {\bibfield  {journal} {\bibinfo  {journal} {Cell Systems}\ }\textbf {\bibinfo
  {volume} {5}},\ \bibinfo {pages} {345} (\bibinfo {year} {2017})}\BibitemShut
  {NoStop}%
\bibitem [{\citenamefont {Hillesland}\ and\ \citenamefont
  {Stahl}(2010)}]{HilleslandStahl_PNAS2010}%
  \BibitemOpen
  \bibfield  {author} {\bibinfo {author} {\bibfnamefont {K.~L.}\ \bibnamefont
  {Hillesland}}\ and\ \bibinfo {author} {\bibfnamefont {D.~A.}\ \bibnamefont
  {Stahl}},\ }\bibfield  {title} {\bibinfo {title} {Rapid evolution of
  stability and productivity at the origin of a microbial mutualism},\ }\href
  {https://doi.org/10.1073/pnas.0908456107} {\bibfield  {journal} {\bibinfo
  {journal} {Proceedings of the National Academy of Sciences}\ }\textbf
  {\bibinfo {volume} {107}},\ \bibinfo {pages} {2124} (\bibinfo {year}
  {2010})},\ \Eprint
  {https://arxiv.org/abs/https://www.pnas.org/doi/pdf/10.1073/pnas.0908456107}
  {https://www.pnas.org/doi/pdf/10.1073/pnas.0908456107} \BibitemShut {NoStop}%
\bibitem [{\citenamefont {Wintermute}\ and\ \citenamefont
  {Silver}(2010)}]{WintermuteSilver_MolSystBiol2010}%
  \BibitemOpen
  \bibfield  {author} {\bibinfo {author} {\bibfnamefont {E.~H.}\ \bibnamefont
  {Wintermute}}\ and\ \bibinfo {author} {\bibfnamefont {P.~A.}\ \bibnamefont
  {Silver}},\ }\bibfield  {title} {\bibinfo {title} {Emergent cooperation in
  microbial metabolism},\ }\href {https://doi.org/10.1038/msb.2010.66}
  {\bibfield  {journal} {\bibinfo  {journal} {Molecular Systems Biology}\
  }\textbf {\bibinfo {volume} {6}},\ \bibinfo {pages} {MSB201066} (\bibinfo
  {year} {2010})}\BibitemShut {NoStop}%
\bibitem [{\citenamefont {Yamagishi}\ \emph {et~al.}(2021)\citenamefont
  {Yamagishi}, \citenamefont {Saito},\ and\ \citenamefont
  {Kaneko}}]{YamagishiSaitoKaneko2021}%
  \BibitemOpen
  \bibfield  {author} {\bibinfo {author} {\bibfnamefont {J.~F.}\ \bibnamefont
  {Yamagishi}}, \bibinfo {author} {\bibfnamefont {N.}~\bibnamefont {Saito}},\
  and\ \bibinfo {author} {\bibfnamefont {K.}~\bibnamefont {Kaneko}},\
  }\bibfield  {title} {\bibinfo {title} {Adaptation of metabolite leakiness
  leads to symbiotic chemical exchange and to a resilient microbial
  ecosystem},\ }\href {https://doi.org/10.1371/journal.pcbi.1009143} {\bibfield
   {journal} {\bibinfo  {journal} {PLoS Computational Biology}\ }\textbf
  {\bibinfo {volume} {17}},\ \bibinfo {pages} {e1009143} (\bibinfo {year}
  {2021})}\BibitemShut {NoStop}%
\bibitem [{\citenamefont {Clegg}\ and\ \citenamefont
  {Gross}(2025)}]{Clegg-Gross_PNAS2025}%
  \BibitemOpen
  \bibfield  {author} {\bibinfo {author} {\bibfnamefont {T.}~\bibnamefont
  {Clegg}}\ and\ \bibinfo {author} {\bibfnamefont {T.}~\bibnamefont {Gross}},\
  }\bibfield  {title} {\bibinfo {title} {Cross-feeding creates tipping points
  in microbiome diversity},\ }\href {https://doi.org/10.1073/pnas.2425603122}
  {\bibfield  {journal} {\bibinfo  {journal} {Proceedings of the National
  Academy of Sciences}\ }\textbf {\bibinfo {volume} {122}},\ \bibinfo {pages}
  {e2425603122} (\bibinfo {year} {2025})},\ \Eprint
  {https://arxiv.org/abs/https://www.pnas.org/doi/pdf/10.1073/pnas.2425603122}
  {https://www.pnas.org/doi/pdf/10.1073/pnas.2425603122} \BibitemShut {NoStop}%
\bibitem [{\citenamefont {Mallmin}\ \emph {et~al.}(2024)\citenamefont
  {Mallmin}, \citenamefont {Traulsen},\ and\ \citenamefont
  {Monte}}]{MallminMonte_PNAS2024}%
  \BibitemOpen
  \bibfield  {author} {\bibinfo {author} {\bibfnamefont {E.}~\bibnamefont
  {Mallmin}}, \bibinfo {author} {\bibfnamefont {A.}~\bibnamefont {Traulsen}},\
  and\ \bibinfo {author} {\bibfnamefont {S.~D.}\ \bibnamefont {Monte}},\
  }\bibfield  {title} {\bibinfo {title} {Chaotic turnover of rare and abundant
  species in a strongly interacting model community},\ }\href
  {https://doi.org/10.1073/pnas.2312822121} {\bibfield  {journal} {\bibinfo
  {journal} {Proceedings of the National Academy of Sciences}\ }\textbf
  {\bibinfo {volume} {121}},\ \bibinfo {pages} {e2312822121} (\bibinfo {year}
  {2024})},\ \Eprint
  {https://arxiv.org/abs/https://www.pnas.org/doi/pdf/10.1073/pnas.2312822121}
  {https://www.pnas.org/doi/pdf/10.1073/pnas.2312822121} \BibitemShut {NoStop}%
\bibitem [{Note1()}]{Note1}%
  \BibitemOpen
  \bibinfo {note} {The detection limit $\delta $ is set at a certain level
  which is much smaller than the populations of major species and is much
  higher than the imposed lower limit, typically at $\delta = \protect \sqrt
  {\epsilon }$. The result in the following is not essentially changed as long
  as $\epsilon $ is enough small and the $\delta $ satisfy the same condition,
  for example, $\epsilon \in [10^{-12}, 10^{-20}], \ \delta = \protect \sqrt
  {\epsilon }$.}\BibitemShut {Stop}%
\bibitem [{\citenamefont {Motomura}(1932)}]{Motomura1932}%
  \BibitemOpen
  \bibfield  {author} {\bibinfo {author} {\bibfnamefont {I.}~\bibnamefont
  {Motomura}},\ }\bibfield  {title} {\bibinfo {title} {On the statistical
  treatment of communities (in japanese)},\ }\href@noop {} {\bibfield
  {journal} {\bibinfo  {journal} {Zool. Mag. (Tokyo)}\ }\textbf {\bibinfo
  {volume} {44}},\ \bibinfo {pages} {379} (\bibinfo {year} {1932})}\BibitemShut
  {NoStop}%
\bibitem [{\citenamefont {Doi}\ and\ \citenamefont
  {Mori}(2013)}]{DoiMori_Oikos2013}%
  \BibitemOpen
  \bibfield  {author} {\bibinfo {author} {\bibfnamefont {H.}~\bibnamefont
  {Doi}}\ and\ \bibinfo {author} {\bibfnamefont {T.}~\bibnamefont {Mori}},\
  }\bibfield  {title} {\bibinfo {title} {The discovery of species-abundance
  distribution in an ecological community},\ }\href
  {https://doi.org/https://doi.org/10.1111/j.1600-0706.2012.00068.x} {\bibfield
   {journal} {\bibinfo  {journal} {Oikos}\ }\textbf {\bibinfo {volume} {122}},\
  \bibinfo {pages} {179} (\bibinfo {year} {2013})}\BibitemShut {NoStop}%
\bibitem [{\citenamefont {Preston}(1962)}]{Preston1962}%
  \BibitemOpen
  \bibfield  {author} {\bibinfo {author} {\bibfnamefont {F.~W.}\ \bibnamefont
  {Preston}},\ }\bibfield  {title} {\bibinfo {title} {The canonical
  distribution of commonness and rarity: Part i},\ }\href
  {https://doi.org/https://doi.org/10.2307/1931976} {\bibfield  {journal}
  {\bibinfo  {journal} {Ecology}\ }\textbf {\bibinfo {volume} {43}},\ \bibinfo
  {pages} {185} (\bibinfo {year} {1962})}\BibitemShut {NoStop}%
\bibitem [{\citenamefont {Volkov}\ \emph {et~al.}(2003)\citenamefont {Volkov},
  \citenamefont {Banavar}, \citenamefont {Hubbell},\ and\ \citenamefont
  {Maritan}}]{VolkovHubbellMaritan2003}%
  \BibitemOpen
  \bibfield  {author} {\bibinfo {author} {\bibfnamefont {I.}~\bibnamefont
  {Volkov}}, \bibinfo {author} {\bibfnamefont {J.~R.}\ \bibnamefont {Banavar}},
  \bibinfo {author} {\bibfnamefont {S.~P.}\ \bibnamefont {Hubbell}},\ and\
  \bibinfo {author} {\bibfnamefont {A.}~\bibnamefont {Maritan}},\ }\bibfield
  {title} {\bibinfo {title} {Neutral theory and relative species abundance in
  ecology},\ }\href {https://doi.org/10.1038/nature01883} {\bibfield  {journal}
  {\bibinfo  {journal} {Nature}\ }\textbf {\bibinfo {volume} {424}},\ \bibinfo
  {pages} {1035} (\bibinfo {year} {2003})}\BibitemShut {NoStop}%
\bibitem [{\citenamefont {Shoemaker}\ \emph {et~al.}(2017)\citenamefont
  {Shoemaker}, \citenamefont {Locey},\ and\ \citenamefont
  {Lennon}}]{Shoemaker2017}%
  \BibitemOpen
  \bibfield  {author} {\bibinfo {author} {\bibfnamefont {W.~R.}\ \bibnamefont
  {Shoemaker}}, \bibinfo {author} {\bibfnamefont {K.~J.}\ \bibnamefont
  {Locey}},\ and\ \bibinfo {author} {\bibfnamefont {J.~T.}\ \bibnamefont
  {Lennon}},\ }\bibfield  {title} {\bibinfo {title} {A macroecological theory
  of microbial biodiversity},\ }\href {https://doi.org/10.1038/s41559-017-0107}
  {\bibfield  {journal} {\bibinfo  {journal} {Nature Ecology \& Evolution}\
  }\textbf {\bibinfo {volume} {1}},\ \bibinfo {pages} {0107} (\bibinfo {year}
  {2017})}\BibitemShut {NoStop}%
\bibitem [{\citenamefont {Bak}\ and\ \citenamefont
  {Sneppen}(1993)}]{BakSneppen1993}%
  \BibitemOpen
  \bibfield  {author} {\bibinfo {author} {\bibfnamefont {P.}~\bibnamefont
  {Bak}}\ and\ \bibinfo {author} {\bibfnamefont {K.}~\bibnamefont {Sneppen}},\
  }\bibfield  {title} {\bibinfo {title} {Punctuated equilibrium and criticality
  in a simple model of evolution},\ }\href
  {https://doi.org/10.1103/PhysRevLett.71.4083} {\bibfield  {journal} {\bibinfo
   {journal} {Phys. Rev. Lett.}\ }\textbf {\bibinfo {volume} {71}},\ \bibinfo
  {pages} {4083} (\bibinfo {year} {1993})}\BibitemShut {NoStop}%
\bibitem [{\citenamefont {Flyvbjerg}\ \emph {et~al.}(1993)\citenamefont
  {Flyvbjerg}, \citenamefont {Bak},\ and\ \citenamefont
  {Sneppen}}]{Flyvbjerg1993}%
  \BibitemOpen
  \bibfield  {author} {\bibinfo {author} {\bibfnamefont {H.}~\bibnamefont
  {Flyvbjerg}}, \bibinfo {author} {\bibfnamefont {P.}~\bibnamefont {Bak}},\
  and\ \bibinfo {author} {\bibfnamefont {K.}~\bibnamefont {Sneppen}},\
  }\bibfield  {title} {\bibinfo {title} {Mean field theory for a simple model
  of evolution},\ }\href@noop {} {\bibfield  {journal} {\bibinfo  {journal}
  {Phys. Rev. Lett.}\ }\textbf {\bibinfo {volume} {71}},\ \bibinfo {pages}
  {4087} (\bibinfo {year} {1993})}\BibitemShut {NoStop}%
\bibitem [{\citenamefont {Christensen}\ \emph {et~al.}(2002)\citenamefont
  {Christensen}, \citenamefont {di~Collobiano}, \citenamefont {Hall},\ and\
  \citenamefont {Jensen}}]{Christensen2002}%
  \BibitemOpen
  \bibfield  {author} {\bibinfo {author} {\bibfnamefont {K.}~\bibnamefont
  {Christensen}}, \bibinfo {author} {\bibfnamefont {S.~A.}\ \bibnamefont
  {di~Collobiano}}, \bibinfo {author} {\bibfnamefont {M.}~\bibnamefont
  {Hall}},\ and\ \bibinfo {author} {\bibfnamefont {H.~J.}\ \bibnamefont
  {Jensen}},\ }\bibfield  {title} {\bibinfo {title} {Tangled nature: a model of
  evolutionary ecology},\ }\href@noop {} {\bibfield  {journal} {\bibinfo
  {journal} {J. theor. Biol.}\ }\textbf {\bibinfo {volume} {216}},\ \bibinfo
  {pages} {73} (\bibinfo {year} {2002})}\BibitemShut {NoStop}%
\bibitem [{\citenamefont {Murase}\ \emph {et~al.}(2010)\citenamefont {Murase},
  \citenamefont {Shimada}, \citenamefont {Ito},\ and\ \citenamefont
  {Rikvold}}]{MuraseRikvold_JTB2010}%
  \BibitemOpen
  \bibfield  {author} {\bibinfo {author} {\bibfnamefont {Y.}~\bibnamefont
  {Murase}}, \bibinfo {author} {\bibfnamefont {T.}~\bibnamefont {Shimada}},
  \bibinfo {author} {\bibfnamefont {N.}~\bibnamefont {Ito}},\ and\ \bibinfo
  {author} {\bibfnamefont {P.~A.}\ \bibnamefont {Rikvold}},\ }\bibfield
  {title} {\bibinfo {title} {Random walk in genome space: A key ingredient of
  intermittent dynamics of community assembly on evolutionary time scales},\
  }\href {https://doi.org/https://doi.org/10.1016/j.jtbi.2010.03.043}
  {\bibfield  {journal} {\bibinfo  {journal} {Journal of Theoretical Biology}\
  }\textbf {\bibinfo {volume} {264}},\ \bibinfo {pages} {663} (\bibinfo {year}
  {2010})}\BibitemShut {NoStop}%
\bibitem [{\citenamefont {Kaneko}\ and\ \citenamefont
  {Tsuda}(2003)}]{ChaoticItineracy_Chaos2003}%
  \BibitemOpen
  \bibfield  {author} {\bibinfo {author} {\bibfnamefont {K.}~\bibnamefont
  {Kaneko}}\ and\ \bibinfo {author} {\bibfnamefont {I.}~\bibnamefont {Tsuda}},\
  }\bibfield  {title} {\bibinfo {title} {Chaotic itinerancy},\ }\href
  {https://doi.org/10.1063/1.1607783} {\bibfield  {journal} {\bibinfo
  {journal} {Chaos: An Interdisciplinary Journal of Nonlinear Science}\
  }\textbf {\bibinfo {volume} {13}},\ \bibinfo {pages} {926} (\bibinfo {year}
  {2003})}\BibitemShut {NoStop}%
\bibitem [{\citenamefont {Beninc{\`a}}\ \emph {et~al.}(2008)\citenamefont
  {Beninc{\`a}}, \citenamefont {Huisman}, \citenamefont {Heerkloss},
  \citenamefont {J{\"o}hnk}, \citenamefont {Branco}, \citenamefont {Van~Nes},
  \citenamefont {Scheffer},\ and\ \citenamefont
  {Ellner}}]{Srephen_PlanktonChaos_Nature2008}%
  \BibitemOpen
  \bibfield  {author} {\bibinfo {author} {\bibfnamefont {E.}~\bibnamefont
  {Beninc{\`a}}}, \bibinfo {author} {\bibfnamefont {J.}~\bibnamefont
  {Huisman}}, \bibinfo {author} {\bibfnamefont {R.}~\bibnamefont {Heerkloss}},
  \bibinfo {author} {\bibfnamefont {K.~D.}\ \bibnamefont {J{\"o}hnk}}, \bibinfo
  {author} {\bibfnamefont {P.}~\bibnamefont {Branco}}, \bibinfo {author}
  {\bibfnamefont {E.~H.}\ \bibnamefont {Van~Nes}}, \bibinfo {author}
  {\bibfnamefont {M.}~\bibnamefont {Scheffer}},\ and\ \bibinfo {author}
  {\bibfnamefont {S.~P.}\ \bibnamefont {Ellner}},\ }\bibfield  {title}
  {\bibinfo {title} {Chaos in a long-term experiment with a plankton
  community},\ }\href {https://doi.org/10.1038/nature06512} {\bibfield
  {journal} {\bibinfo  {journal} {Nature}\ }\textbf {\bibinfo {volume} {451}},\
  \bibinfo {pages} {822} (\bibinfo {year} {2008})}\BibitemShut {NoStop}%
\bibitem [{\citenamefont {Rogers}\ \emph {et~al.}(2022)\citenamefont {Rogers},
  \citenamefont {Johnson},\ and\ \citenamefont
  {Munch}}]{Sthephan_UbiquitusChaos_NatEcoEvo2022}%
  \BibitemOpen
  \bibfield  {author} {\bibinfo {author} {\bibfnamefont {T.~L.}\ \bibnamefont
  {Rogers}}, \bibinfo {author} {\bibfnamefont {B.~J.}\ \bibnamefont
  {Johnson}},\ and\ \bibinfo {author} {\bibfnamefont {S.~B.}\ \bibnamefont
  {Munch}},\ }\bibfield  {title} {\bibinfo {title} {Chaos is not rare in
  natural ecosystems},\ }\href {https://doi.org/10.1038/s41559-022-01787-y}
  {\bibfield  {journal} {\bibinfo  {journal} {Nature Ecology \& Evolution}\
  }\textbf {\bibinfo {volume} {6}},\ \bibinfo {pages} {1105} (\bibinfo {year}
  {2022})}\BibitemShut {NoStop}%
\bibitem [{\citenamefont {Shimada}\ \emph {et~al.}(2024)\citenamefont
  {Shimada}, \citenamefont {Mise}, \citenamefont {Morino},\ and\ \citenamefont
  {Otsuka}}]{Ecosoil_Shimada2024}%
  \BibitemOpen
  \bibfield  {author} {\bibinfo {author} {\bibfnamefont {T.}~\bibnamefont
  {Shimada}}, \bibinfo {author} {\bibfnamefont {K.}~\bibnamefont {Mise}},
  \bibinfo {author} {\bibfnamefont {K.}~\bibnamefont {Morino}},\ and\ \bibinfo
  {author} {\bibfnamefont {S.}~\bibnamefont {Otsuka}},\ }\href
  {https://arxiv.org/abs/2409.03372} {\bibinfo {title} {Simple measures to
  capture the robustness and the plasticity of soil microbial communities}}
  (\bibinfo {year} {2024}),\ \Eprint {https://arxiv.org/abs/2409.03372}
  {arXiv:2409.03372 [q-bio.PE]} \BibitemShut {NoStop}%
\bibitem [{\citenamefont {Arnoulx~de Pirey}\ and\ \citenamefont
  {Bunin}(2024)}]{PireyBunin_PRX2024}%
  \BibitemOpen
  \bibfield  {author} {\bibinfo {author} {\bibfnamefont {T.}~\bibnamefont
  {Arnoulx~de Pirey}}\ and\ \bibinfo {author} {\bibfnamefont {G.}~\bibnamefont
  {Bunin}},\ }\bibfield  {title} {\bibinfo {title} {Many-species ecological
  fluctuations as a jump process from the brink of extinction},\ }\href
  {https://doi.org/10.1103/PhysRevX.14.011037} {\bibfield  {journal} {\bibinfo
  {journal} {Phys. Rev. X}\ }\textbf {\bibinfo {volume} {14}},\ \bibinfo
  {pages} {011037} (\bibinfo {year} {2024})}\BibitemShut {NoStop}%
\bibitem [{\citenamefont {Huisman}\ and\ \citenamefont
  {Weissing}(1999)}]{Huisman_DynamicalBreakingOfGause_Nature1999}%
  \BibitemOpen
  \bibfield  {author} {\bibinfo {author} {\bibfnamefont {J.}~\bibnamefont
  {Huisman}}\ and\ \bibinfo {author} {\bibfnamefont {F.~J.}\ \bibnamefont
  {Weissing}},\ }\bibfield  {title} {\bibinfo {title} {Biodiversity of plankton
  by species oscillations and chaos},\ }\href {https://doi.org/10.1038/46540}
  {\bibfield  {journal} {\bibinfo  {journal} {Nature}\ }\textbf {\bibinfo
  {volume} {402}},\ \bibinfo {pages} {407} (\bibinfo {year}
  {1999})}\BibitemShut {NoStop}%
\bibitem [{\citenamefont {Shimada}(2014)}]{Takashi_EOS_SREP2014}%
  \BibitemOpen
  \bibfield  {author} {\bibinfo {author} {\bibfnamefont {T.}~\bibnamefont
  {Shimada}},\ }\bibfield  {title} {\bibinfo {title} {A universal transition in
  the robustness of evolving open systems},\ }\href@noop {} {\bibfield
  {journal} {\bibinfo  {journal} {Scientific Reports}\ }\textbf {\bibinfo
  {volume} {4}},\ \bibinfo {pages} {4082} (\bibinfo {year} {2014})}\BibitemShut
  {NoStop}%
\bibitem [{\citenamefont {Ogushi}\ \emph {et~al.}(2017)\citenamefont {Ogushi},
  \citenamefont {Kert{\'e}sz}, \citenamefont {Kaski},\ and\ \citenamefont
  {Shimada}}]{Fumiko_EOS_SREP2017}%
  \BibitemOpen
  \bibfield  {author} {\bibinfo {author} {\bibfnamefont {F.}~\bibnamefont
  {Ogushi}}, \bibinfo {author} {\bibfnamefont {J.}~\bibnamefont {Kert{\'e}sz}},
  \bibinfo {author} {\bibfnamefont {K.}~\bibnamefont {Kaski}},\ and\ \bibinfo
  {author} {\bibfnamefont {T.}~\bibnamefont {Shimada}},\ }\bibfield  {title}
  {\bibinfo {title} {Enhanced robustness of evolving open systems by the
  bidirectionality of interactions between elements},\ }\href
  {https://doi.org/10.1038/s41598-017-07283-9} {\bibfield  {journal} {\bibinfo
  {journal} {Scientific Reports}\ }\textbf {\bibinfo {volume} {7}},\ \bibinfo
  {pages} {6978} (\bibinfo {year} {2017})}\BibitemShut {NoStop}%
\bibitem [{\citenamefont {Ogushi}\ \emph {et~al.}(2019)\citenamefont {Ogushi},
  \citenamefont {Kert{\'e}sz}, \citenamefont {Kaski},\ and\ \citenamefont
  {Shimada}}]{Fumiko_EOS_RoySocOpenSci2019}%
  \BibitemOpen
  \bibfield  {author} {\bibinfo {author} {\bibfnamefont {F.}~\bibnamefont
  {Ogushi}}, \bibinfo {author} {\bibfnamefont {J.}~\bibnamefont {Kert{\'e}sz}},
  \bibinfo {author} {\bibfnamefont {K.}~\bibnamefont {Kaski}},\ and\ \bibinfo
  {author} {\bibfnamefont {T.}~\bibnamefont {Shimada}},\ }\bibfield  {title}
  {\bibinfo {title} {Temporal inactivation enhances robustness in an evolving
  system},\ }\href {https://doi.org/10.1098/rsos.181471} {\bibfield  {journal}
  {\bibinfo  {journal} {Royal Society Open Science}\ }\textbf {\bibinfo
  {volume} {6}},\ \bibinfo {pages} {181471} (\bibinfo {year} {2019})},\ \Eprint
  {https://arxiv.org/abs/https://royalsocietypublishing.org/rsos/article-pdf/doi/10.1098/rsos.181471/972567/rsos.181471.pdf}
  {https://royalsocietypublishing.org/rsos/article-pdf/doi/10.1098/rsos.181471/972567/rsos.181471.pdf}
  \BibitemShut {NoStop}%
\end{thebibliography}%
%%%%%%%%%%%%%%%%%%%%%%%%%%%%%%%%%%%%%%%%%%%%%%%%%%%%%%%%%%%%%%%%%%%%%%%%%%%%
%%%%%%%%%%%%%%%%%%%%%%%%%%%%%%%%%%%%%%%%%%%%%%%%%%%%%%%%%%%%%%%%%%%%%%%%%%%%
%%%%%%%%%%%%%%%%%%%%%%%%%%%%%%%%%%%%%%%%%%%%%%%%%%%%%%%%%%%%%%%%%%%%%%%%%%%%
%%%%%%%%%%%%%%%%%%%%%%%%%%%%%%%%%%%%%%%%%%%%%%%%%%%%%%%%%%%%%%%%%%%%%%%%%%%%
\end{document}